\documentclass[aps,prb,pdflatex,reprint,showpacs,superscriptaddress]{revtex4-2}
\usepackage{amsmath}
\usepackage{amssymb}
\usepackage{amsthm}
\usepackage{bbm}
\usepackage{graphicx}
\usepackage{hyperref}
\usepackage{physics}
\usepackage{systeme}
\usepackage{times}
\usepackage{epstopdf}
\usepackage{float}
\usepackage{xcolor}

\hypersetup{
  colorlinks=true,linkcolor=blue,citecolor=blue,
  filecolor=blue,urlcolor=blue,breaklinks=true
}

\begin{document}

\title{On the Dirac oscillator in a spinning cosmic string spacetime in external magnetic fields: Investigation of energy spectrum and connection with the condensed matter physics}
\author{M\'{a}rcio M. Cunha}
\email{marciomc05@gmail.com}
\affiliation{
        Departamento de F\'{i}sica,
        Universidade Federal do Maranh\~{a}o,
        65085-580, S\~{a}o Lu\'{i}s, Maranh\~{a}o, Brazil
      }
\author{Henrique S. Dias}
\email{henrique.dias1915@gmail.com}
\affiliation{
        Departamento de F\'{i}sica,
        Universidade Federal do Maranh\~{a}o,
        65085-580, S\~{a}o Lu\'{i}s, Maranh\~{a}o, Brazil
      }
\author{Edilberto O. Silva}
\email{edilbertoo@gmail.com}
\affiliation{
        Departamento de F\'{i}sica,
        Universidade Federal do Maranh\~{a}o,
        65085-580, S\~{a}o Lu\'{i}s, Maranh\~{a}o, Brazil
      }
\date{\today }
\begin{abstract}
In this article, we study topological and noninertial effects on the motion of the two-dimensional Dirac oscillator in the presence of a uniform magnetic field and the Aharonov-Bohm potential. We obtain the Dirac equation that describes the model. Expressions for the wave functions and energy spectrum are derived. The energy spectrum of the oscillator as a function of the various physical parameters involved in the problem is rigorously studied. We estimate the phenomenological energy scale of the model based on upper bounds for string tension, as well as values for the angular momentum $J$ and the intergalactic magnetic fields found in literature. Finally, we present some analogies between our results and others of the condensed matter physics. 
\end{abstract}

\maketitle

\section{Introduction}
\label{intro}

The harmonic oscillator model plays a crucial role in the comprehension of many phenomena in both classical mechanics and nonrelativistic quantum mechanics.
A relevant question involving this model consists in thinking about how to make its appropriate description in the scenario of relativistic quantum mechanics.
Initially, It{\^o}, Mori, and Carrieri, by using the minimal substitution, introduced a linear dependence in the radial coordinate into the Dirac equation \cite{ito1967example}. They showed it leads to a harmonic oscillator with strong spin-orbit coupling.
Subsequently, Moshinsky and Szczepaniak revived the interest on this issue. 
By employing the same minimal substitution procedure, they named this system as the Dirac Oscillator, because in the nonrelativistic limit it becomes the usual quantum harmonic oscillator containing a spin-orbit term \cite{Moshinsky_1989}. Following this same idea, the full energy spectrum of the Dirac oscillator was also obtained in connection with the supersymmetry \cite{PhysRevLett.64.1643}. A similar strategy was adopted in the context of the Klein-Gordon equation, originating the model known as Klein-Gordon oscillator \cite{bruce1993klein}.

After these initial steps, several fundamental features of the Dirac oscillator were examined. For instance, the covariance, CPT properties, and the Foldy-Wouthuysen transformation were studied \cite{Moreno_1989_cpt_fw}. Other important contributions such as aspects regarding the Lie algebra involved in the model \cite{Quesne_1990}, shift operators \cite{Lange_1991shift} and algebraic properties \cite{Lange_1991algebraic} were examined. The Dirac oscillator was also resolved considering two spatial dimensions \cite{PhysRevA.49.586,doi:10.1142/S0217732304014719}. Years later, this same problem was reexamined and it was shown that the degeneracy of the energy spectrum can occur for all possible values of the quantum number $m$ \cite{EPL.2014.108.30003}.
The construction of a coherent state for the Dirac oscillator also it was studied  \cite{doi:10.1139/p96-018}. A physical interpretation was provided and the Lorentz covariance of the interaction term of the Dirac oscillator was examined \cite{Martinez_y_Romero_1995}.  The completeness of its eigenfunctions it was proved in \cite{Szmytkowski_2001}, and the matrix elements of physical quantities were obtained \cite{Ning_2004_matrix}.
The Dirac oscillator was also studied in a noncommutative spacetime \cite{Mirza_2004_kg_dirac}.
In the context of recent contributions,  the Dirac oscillator was analyzed taking into account a scenario with position-dependent mass \cite{Ho_Roy_2019_EPL} and the inclusion of time-reversal symmetry \cite{IWAI20191389}.

The connection of the Dirac oscillator with accessible experimental models also has attracted attention in the literature. An example of this is a comparison between the Jaynes-Cummings model and the dynamics of the Dirac oscillator \cite{Rozmej_1999_jc_model}. It was showed that the spin-orbit coupling can produce entangled degrees of freedom in the Dirac oscillator. Also, an ion-trap experimental proposal it was reported \cite{PhysRevA.76.041801}. In \cite{Longhi:10}, it was presented a classic wave optics analog of a one-dimensional Dirac oscillator. 
A study focused on the Zitterbewegung behavior of Dirac oscillator it was reported in \cite{wang2012zitterbewegung}.
An experimental realization of the one-dimensional Dirac oscillator was implemented by using a microwave system \cite{PhysRevLett.111.170405}.
A connection between the Dirac oscillator and graphene physics also it was established by using the approach of effective mass \cite{Boumali_2015_graphene_DO}.

In Quantum Mechanics, we often are interested in studying the effect of electromagnetic potentials on the quantum dynamics of a given system. It has been done in the context of the Dirac oscillator. The two-dimensional Dirac oscillator in the presence of a constant magnetic field \cite{villalba_2001energy_magnetic} and the Aharonov-Bohm potential were investigated \cite{FERKOUS200421}. A path-integral formulation for the problem of a Dirac oscillator in the presence of a constant magnetic field was presented in \cite{haouat2007_exact_green}.
 The construction of coherent states for the $(2+1)$-dimensional Dirac oscillator coupled to an external field also was considered \cite{Ojeda_Guill_n_2015}. In \cite{MANDAL20101021}, it was established a connection between the Dirac oscillator and the Anti-Jaynes-Cummings model and Landau levels are obtained explicitly. The problem of a three-dimensional Dirac oscillator subjected to Aharonov-Bohm and magnetic Monopole potentials it was addressed in \cite{alhaidari2005three}. The appearance of a relativistic quantum phase transition was reported to the Dirac oscillator interacting with a magnetic field in the case of a usual oscillator \cite{PhysRevA.77.063815} and the noncommutative oscillator \cite{PhysRevA.90.042111}. Thermal properties of the Dirac oscillator in this context also were investigated \cite{boumali_2013_thermal_EPJ_B,Frassino_Roy_2020_JPA}. In \cite{OLIVEIRA20191_AB_Coulomb}, the Dirac oscillator was analyzed in the presence of a magnetic field in an Aharonov-Bohm-Coulomb system.
 
Another pertinent aspect of studying quantum systems is related to incorporate the influence of geometry in physical properties of interest. In this context, we can be interested in analyzing how the presence of curvature, for example, can affect a given system \cite{DU201628}. Besides, we can explore the quantum dynamics of a system when it is immersed in a spacetime having a topological defect \cite{medeiros2012relativistic}. It is also a relevant issue since topological defects can take place in many physical systems, covering research areas such as Cosmology \cite{doi:10.1142/S0217751X9400090X} and Condensed Matter Physics \cite{CORTIJO2007293}.
The Dirac oscillator in topological defects backgrounds has been investigated in several scenarios. For instance, the energy spectrum \cite{PhysRevA.84.032109} and coherent states \cite{salazar_2019_algebraic} for the case of a cosmic string spacetime were investigated.
The Dirac oscillator interacting with an Aharonov-Casher system in the presence of topological defects was studied in \cite{BAKKE2013489}.
The cosmic string spacetime was considered to the formulation of a generalized Dirac oscillator \cite{deng2018_generalized_DO_cosmic}, and also in the case of the Dirac oscillator in the context of spin and pseudospin symmetries \cite{EPJC.2019.79.596}.

Besides the influence of electromagnetic interactions and topological defects, noninertial effects are associated with important contributions to the dynamics of a system. More specifically, such effects can be related to Hall quantization \cite{Fischer_2001_EPL}, geometric phases \cite{PhysRevB.68.195421}, spin currents \cite{DAYI2018143} and modifications in the energy spectrum of a system \cite{shen2005_aharonov_EPJD}. Focusing our attention on the Dirac oscillator again, it was analyzed in a rotating frame of reference in \cite{STRANGE20163465}.
Also, the combined influence of topological defect and noninertial effects in the Dirac oscillator has been investigated.
In \cite{EPJC.2019.79.311}, for instance, it was considered a spinning cosmic string spacetime, and the eigenfunctions and energy levels were obtained. Topological and noninertial contributions also were analyzed in the case of the Aharonov–Casher effect \cite{EPJC_oliveira2019topological}.

Then, examining simultaneously electromagnetic, topological, and noninertial effects on the motion of the Dirac oscillator it is also a valid discussion and is a generalization of other studies in this context in the literature. 
From this motivation, the aim of the present manuscript consists in describe the problem of a Dirac oscillator in the presence of a topological defect, rotation, and external magnetic fields. More precisely, we solve the problem of the Dirac oscillator in the spinning cosmic string background taking into account the presence of the Aharonov-Bohm potential and a uniform magnetic field. 

The organization of the paper is as follows. In Sec. \ref{SecII}, we make a brief review of the elements necessary to write the Dirac equation in curved time. We define the magnetic field configuration and the substitution that allows the inclusion of the Dirac oscillator in the model. Using an appropriate ansatz, we obtain the motion equation describing the Dirac oscillator in the spinning cosmic string background in the presence of Aharonov-Bohm potential and uniform magnetic field. In Sec. \ref{SecIII}, we derived the corresponding radial equation of motion. We solve this equation and obtain the eigenfunctions and the energy spectrum of the oscillator. We make a rigorous inspection to understand how the physical parameters associated with the topological defect, rotation, magnetic field, and Aharanov-Bohm potential affect the energy spectrum of the Dirac oscillator. To make our realization clearer, we make several energy sketches as a function of the parameters involved.
We finish Sec. \ref{SecIII} by presenting some phenomenological estimates for the energy scale of our model.
In Sec. \ref{SecIV}, we present some similarities between our results and a model for Topological Insulators in Condensed Matter Physics.
The paper is summarized and concluded in Sec. \ref{SecV}. In this article, we employ natural units, $\hbar = c = G = 1$.

\section{Dirac equation in the spinning cosmic string spacetime}\label{SecII}

The goal of this section is to obtain the equation that describes the motion of the Dirac oscillator in the spinning cosmic string spacetime in the presence of a uniform field and the Aharonov-Bohm potential. Before this, we present some elements necessary for the construction of such an equation in curved space-time. Let us begin with the  Dirac equation in a generic curved spacetime
\begin{equation}
\left[ i\gamma ^{\mu }\left( x\right) \left( \partial _{\mu }+\Gamma _{\mu
}\left( x\right) +ieA_{\mu }(x)\right) -M\right] \Psi \left( x\right) =0,
\label{diracsc}
\end{equation}
where $\gamma ^{\mu }\left( x\right)$ are the Dirac matrices in the curved space and $M$ represents the particle mass. The Dirac matrices are related to their counterparts in the Minkowski spacetime, $\gamma^{a}$, by the relation
\begin{equation}
\gamma ^{\mu }\left( x\right)=e_{a}^{\mu }\left( x\right) \gamma ^{a},
\label{gmatrices}
\end{equation}
being $e_{a}^{\mu }(x)$ the tetrad fields. Explicitly, the Dirac matrices in the flat space are given by
\begin{equation}
\gamma ^{a}=\left( \gamma ^{0},\gamma ^{i}\right),
\end{equation}
with
\begin{equation}
   \gamma ^{0}=\left(
\begin{array}{cc}
I & 0 \\
0 & -I
\end{array}
\right),\;\gamma ^{i}=\left(
\begin{array}{cc}
0 & \sigma ^{i} \\
-\sigma ^{i} & 0
\end{array}
\right),
\end{equation}
where $I$ is a $2 \times 2$ unit matrix and $\sigma ^{i}=\left( \sigma ^{x},\sigma ^{y},\sigma ^{z}\right)$ are the standard Pauli matrices.
The matrices (\ref{gmatrices}) obey the following algebraic property:
\begin{equation}
\left\{ \gamma ^{\mu }\left( x\right) ,\gamma ^{\nu }\left( x\right)
\right\} =2g^{\mu \nu }\left( x\right).
\end{equation}
The tetrad field satisfies the following relations:
\begin{align}
&e_{\mu }^{a}\left( x\right) e_{\nu}^{b}\left(x\right) \eta_{ab}=g_{\mu \nu }\left( x\right),\\
&e_{\mu}^{a}\left(x\right) e_{b
}^{\mu}\left(x\right)=\delta_{b}^{a},\\
&e_{a}^{\mu}\left(x\right)
e_{\nu}^{a}\left(x\right) =\delta_{\nu}^{\mu},
\end{align}
where $\eta_{ab}$ and $g_{\mu \nu }$ are the metric tensor for the Minkowski space and the curved space, respectively. Here, the Latin letters refer to Minkowski indices while the Greek letters are related to the curved coordinates.
Finally, the quantity $\Gamma _{\mu }\left( x\right)$  in Eq. (\ref{diracsc}) represents the
spin affine connection, which has the form
\begin{equation}
\Gamma _{\mu }\left( x\right) =\frac{1}{4}\gamma ^{a}\gamma ^{b}e_{a}^{\nu
}\left( x\right) \left[ \partial _{\mu }e_{b\nu }\left( x\right) -\Gamma
_{\mu \nu }^{\sigma }e_{b\sigma }\left( x\right) \right], \label{affine}
\end{equation}
with $\Gamma_{\mu \nu}^{\sigma}$ being the Christoffel symbols of the second kind. We have all the elements we need to write the Dirac equation in curved spacetime. Now, let us specialize to the spinning cosmic string spacetime, whose line element (written in cylindrical coordinates) is given by
\begin{equation}
ds^{2}=\left( dt+ad\varphi \right) ^{2}-dr^{2}-\alpha ^{2}r^{2}d\varphi
^{2}-dz^{2},  \label{metric}
\end{equation}
where $-\infty <z<\infty $, $r\geqslant 0$ and $0\leqslant \varphi \leqslant
2\pi$. Also, the parameter $\alpha $ is associated with the linear mass density $\mu$ following the relation
$\alpha =1-4\mu$, and it is defined in the range $(0,1]$. The presence of rotation is characterized by the parameter $a=4J$, and $J$ is the angular momenta of the cosmic string.
This spacetime admits closed timelike curves \cite{PhysRevD.45.3528} if $ r< |a|/\alpha$ . Then, here, we are interested only in regions such that $r>|a|/\alpha.$
The next quantity we need to define is the tetrad field. We use the tetrad basis and its inverse given by \cite{EPJC.2019.79.311}
\begin{eqnarray}
e_{\mu }^{a}\left( x\right)  &=&\left(
\begin{array}{cccc}
1 & 0 & a & 0 \\
0 & \cos \varphi  & -r\alpha \sin \varphi  & 0 \\
0 & \sin \varphi  & r\alpha \cos \varphi  & 0 \\
0 & 0 & 0 & 1
\end{array}
\right) , \\
e_{a}^{\mu }\left( x\right)  &=&\left(
\begin{array}{cccc}
1 & \frac{a\sin \varphi }{r\alpha } & -\frac{a\cos \varphi }{r\alpha } & 0
\\
0 & \cos \varphi  & \sin \varphi  & 0 \\
0 & -\frac{\sin \varphi }{r\alpha } & \frac{\cos \varphi }{r\alpha } & 0 \\
0 & 0 & 0 & 1
\end{array}
\right).\label{trdi}
\end{eqnarray}
By using these tetrad fields, it can be demonstrated the unique non-vanishing contribution to the affine connection is found to be
\begin{equation}
\Gamma _{\mu }=\left( 0,0,\Gamma _{\varphi },0\right),
\end{equation}
with
\begin{equation}
\Gamma _{\varphi }=\frac{i}{2}\left( 1-\alpha \right) \Sigma^{z},
\end{equation}
and
\begin{equation}
\Sigma ^{z}=\left(
\begin{array}{cc}
\sigma ^{z} & 0 \\
0 & \sigma ^{z}
\end{array}
\right), \;\; \sigma ^{z} =\left(
\begin{array}{cc}
1 & 0 \\
0 & -1%
\end{array}
\right). \label{sz}
\end{equation}
Note that the spin connection $\Gamma_{\mu }$ is identically null if $\alpha=1$. The tetrad basis $e_{a}^{\mu}(x)$ must be used to obtain the Dirac matrices (\ref{gmatrices}) in the spacetime of the spinning cosmic string. Such matrices are found to be
\begin{align}
\gamma ^{t} &=e_{0}^{t}\gamma ^{0}=\gamma ^{0}-a\gamma ^{\varphi }\notag\\&=\gamma
^{0}-\frac{a}{\alpha r}\left( -\gamma ^{1}\sin \varphi +\gamma ^{2}\cos
\varphi \right) ,  \label{ga0} \\
\gamma ^{r} &=e_{a}^{r}\gamma ^{a}=e_{0}^{r}\gamma ^{0}+e_{1}^{r}\gamma
^{1}+e_{2}^{r}\gamma ^{2}\notag \\&=\gamma ^{1}\cos \varphi +\gamma ^{2}\sin \varphi ,\label{gar}
\\
\gamma ^{\varphi } &=e_{a}^{\varphi }\gamma ^{a}=e_{0}^{\varphi }\gamma
^{0}+e_{1}^{\varphi }\gamma ^{1}+e_{2}^{\varphi }\gamma ^{2}\notag \\&=\frac{1}{\alpha
r}\left( -\gamma ^{1}\sin \varphi +\gamma ^{2}\cos \varphi \right),\label{gaphi}  \\
\gamma ^{z}&=e_{0}^{z}\gamma ^{0}=\gamma ^{3}.  \label{gaz}
\end{align}
We can also define the $\alpha ^{i}(x)$ matrices, which can be written in terms
of Pauli's matrices as
\begin{equation}
\alpha ^{i}\left( x\right) =e_{a}^{i}\left( x\right) \left(
\begin{array}{cc}
0 & \sigma ^{a} \\
\sigma ^{a} & 0
\end{array}
\right) =\left(
\begin{array}{cc}
0 & \sigma ^{i}(x) \\
\sigma ^{i}(x) & 0
\end{array}
\right) ,  \label{alphad}
\end{equation}
where $\sigma ^{i}=\left( \sigma ^{r},\sigma ^{\varphi },\sigma ^{z}\right) $
are the Pauli matrices in cylindrical coordinates written on the tetrad basis $e_{\mu }^{a}\left( x\right) $.

In this representation, the $\gamma^{\mu}\left(x\right)$ and $\sigma^{i}(x)$ matrices are explicitly written as
\begin{align}
&\gamma ^{t} =\left(
\begin{array}{cc}
I & -a\,\sigma ^{\varphi } \\
a\,\sigma ^{\varphi } & -I
\end{array}
\right),\;\;
\gamma ^{r}=\left(
\begin{array}{cc}
0 & \sigma ^{r} \\
-\sigma ^{r} & 0
\end{array}
\right), \label{grr}\\ &\gamma ^{\varphi }=\left(
\begin{array}{cc}
0 & \sigma ^{\varphi } \\
-\sigma ^{\varphi } & 0
\end{array}
\right),\;\; \gamma ^{z}=\left(
\begin{array}{cc}
0 & \sigma ^{z} \\
-\sigma ^{z} & 0
\end{array}
\right), \label{grphi}
\end{align}
and
\begin{equation}
\sigma^{r} =\left(
\begin{array}{cc}
0 & e^{-i\varphi} \\
e^{+i\varphi } & 0
\end{array}
\right), \;\;\sigma ^{\varphi }=\frac{1}{r\alpha }\left(
\begin{array}{cc}
0 & -ie^{-i\varphi } \\
ie^{+i\varphi } & 0
\end{array}
\right).\label{sigmat}
\end{equation}
Now, we need to include the interactions to the model. First, we introduce the Dirac oscillator \cite{Moshinsky_1989}. The Dirac oscillator is an important model in relativistic quantum mechanics because it can be solved exactly \cite{Martinez_y_Romero_1995,PhysRevA.49.586}. The coupling that describes the oscillator makes the Dirac equation linear in both the momenta and the spatial coordinates. The oscillator is included in Dirac's equation by replacing
\begin{equation}
\mathbf{p\rightarrow p}-iM\omega \gamma ^{0}\mathbf{r}=-i\mathbf{
\nabla }-iM\omega \gamma ^{0}\mathbf{r}.
\end{equation}
Although this substitution is non hermitian, the hermiticity of the complete Hamiltonian is guaranteed by the presence of the matrix $\alpha^{i}(x)$. To describe the Dirac oscillator in the presence of a magnetic field and Aharonov-Bohm potential, we consider the minimal substitution procedure. The particle is subject to the following field configuration:
\begin{equation}
B=B_{z,1}+B_{z,2},
\end{equation}
being
\begin{equation}
B_{1,z}=B,  \;\;\; B_{z,2}=\frac{\phi}{e} \frac{\delta (r)}{\alpha r} \label{fields}
  \end{equation}
the contributions due the uniform field and the Aharonov-Bohm effect, respectively. The above field configuration is the result of the superposition of vector potentials (in the Coulomb gauge) given by
\begin{equation}
\mathbf{A}=\left( 0, A_{\varphi }, 0\right),\;\; \text{with} \;\; \mathbf{\nabla} \cdot \mathbf{A}=0,  \label{vectorA}
\end{equation}
with
\begin{align}
A_{\varphi } &=-\left(A_{\varphi ,1}+A_{\varphi,2}\right) , \label{potential} \\
A_{\varphi ,1} &=\frac{1}{2}\alpha Br^{2},\;\;\; A_{\varphi ,2}=\frac{\phi}{e},\label{potentialv}
\end{align}
where $B$ is the amplitude of the uniform magnetic field, $\phi =\Phi /\Phi _{0}$ is related to the Aharonov-Bohm flux, which $\Phi$ represents the
magnetic flux and $\Phi _{0}=2\pi /e$ is the quantum of magnetic flux.
We are interested in studying the quantum particle motion in the region $r >0$. Then, in this case, the particle does not interact directly with the field $B_{z,2}$, despite the fact it will suffer the influence of the Aharonov-Bohm flux.
As a consequence, we deal only with regular solutions to the wave functions.
Also, it is important to note that the system has translational invariance in the $z$-direction, in such a way we can exclude the $z$ degree of freedom by imposing $p_z=z=0$, resulting in a planar dynamics \cite{PRL.1990.64.503,PRD.1994.50.7715,PRD.2012.85.041701,AoP.2013.339.510}.
Thus, using the stationary solution of energy $E$,
\begin{align}
    \Psi \left( r,\varphi \right)=e^{-iEt}\left(
\begin{array}{c}
\psi _{1}\left( r,\varphi \right)  \\
\psi _{2}\left( r,\varphi \right)
\end{array}
\right),\label{S1}
\end{align}
the Dirac equation (\ref{diracsc}) assumes the following form:
\begin{align}
& \left( E-M\right) \psi_{1}+\sigma ^{r}\left( i\partial_{r}-iM\omega
r\right) \psi_{2}  \notag \\
&+\sigma^{\varphi}\left( i\partial _{\varphi }-eA_{\varphi }-aE-\frac{\sigma^z}{2}%
\left( 1-\alpha \right) \right) \psi _{2}=0, \label{Eqpsi}\\
& \left( E+M\right) \psi _{2}+\sigma ^{r}\left( i\partial _{i}+iM\omega
r\right) \psi_{1} \notag \\
&+\sigma ^{\varphi }\left( i\partial _{\varphi }-eA_{\varphi }-aE-\frac{\sigma^z}{2}%
\left( 1-\alpha \right) \right) \psi _{1}=0. \label{Eqchi}
\end{align}
From these equations, we can derive both second-order equations for the components $\psi_{1}$ and $\psi_{2}$ of the spinor. Here, for reasons that will be clarified later, we consider only the upper component $\psi_{1}$. This is accomplished in the next section.

\section{Solution of the equation of motion}\label{SecIII}

In this section, we shall solve the second-order equation for $\psi_{1}$ derived from Eqs. (\ref{Eqpsi}) and (\ref{Eqchi}) to obtain the wave functions and energy eigenvalues and then make a detailed analysis of the energy spectrum as well as its physical implications in connection with other physical systems. By solving (\ref{Eqchi}) for $\psi_{2}$, we obtain
\begin{align}
&\psi _{2}=-\frac{i}{E+M}\notag \\&\times\Bigg( \sigma ^{i}\left( \partial
_{i}+ieA_{i}\right) +iaE\sigma ^{\varphi }+M\omega r\sigma ^{r}-\frac{\left( 1-\alpha \right) \sigma ^{r}}{
2\alpha r}\Bigg) \psi _{1}. \label{cpsi2}
\end{align}
Replacing (\ref{cpsi2}) in (\ref{Eqpsi}), we find
\begin{align}
&\left( \sigma ^{i}\left( \partial _{i}+ieA_{i}\right)
+iaE\sigma ^{\varphi }-M\omega r\sigma ^{r}-\frac{\left( 1-\alpha \right)
\sigma ^{r}}{2\alpha r}\right)   \notag \\
& \times \left( \sigma ^{j}\left( \partial _{j}+ieA_{j}\right) +iaE\sigma
^{\varphi }+M\omega r\sigma ^{r}-\frac{\left( 1-\alpha \right) \sigma ^{r}}{%
2\alpha r}\right)\psi _{1} \notag \\ &+\left( E^{2}-M^{2}\right) \psi_{1}=0.\label{secondoe}
\end{align}
The first term of this equation can be developed using the commutation relations between the matrices $\sigma^{r}, \sigma^{\varphi}$ and $\sigma^{z}$. After some algebraic manipulations, Eq. (\ref{secondoe}) results
\begin{align}
& \left( E^{2}-M^{2}\right) \psi _{1}+\partial _{r}^{2}\psi _{1}+\frac{1}{r}
\partial _{r}\psi _{1}+\frac{1}{\alpha ^{2}r^{2}}\partial _{\varphi
}^{2}\psi _{1}  \notag \\
& -\frac{1}{\alpha ^{2}r^{2}}e^{2}\left( A_{\varphi ,1}+A_{\varphi,2}\right) ^{2}\psi _{1}-\frac{1}{4\alpha ^{2}r^{2}}\left( 1-\alpha \right)^{2}\psi _{1}  \notag \\& -\frac{1}{\alpha ^{2}r^{2}}a^{2}E^{2}\psi _{1}+\frac{1}{\alpha ^{2}r^{2}}%
2ie\left( A_{\varphi ,1}+A_{\varphi ,2}\right) \partial _{\varphi }\psi _{1}\notag \\
& -2\frac{1}{\alpha r}\frac{\left( 1-\alpha \right) }{2}\frac{1}{\alpha r}
e\left( A_{\varphi ,1}+A_{\varphi ,2}\right) \sigma ^{z}\psi _{1}+\frac{2aE}{
\alpha ^{2}r^{2}}i\partial_{\varphi} \psi_{1}  \notag \\
& +\frac{2}{\alpha ^{2}r^{2}}\frac{1}{2}\left( 1-\alpha \right)\sigma^{z} i\partial_{\varphi}\psi_{1} -\frac{1}{\alpha r}\sigma ^{z}e\left[\partial _{r}\left( A_{\varphi ,1}+A_{\varphi ,2}\right) \right] \psi _{1}
\notag \\
& -\frac{2aE}{\alpha ^{2}r^{2}}e\left( A_{\varphi ,1}+A_{\varphi ,2}\right)\psi _{1}-\frac{2aE}{\alpha ^{2}r^{2}}\frac{1}{2}\left( 1-\alpha \right)\sigma^{z}\psi_{1}\notag \\
& +2M\omega \psi _{1}-2M\omega \frac{1}{\alpha}\sigma ^{z}i\partial_{\varphi}\psi_{1}+2M\omega \frac{1}{\alpha}\sigma^{z}eA_{\varphi }\psi_{1}  \notag \\
& +M\omega \frac{1}{\alpha }aE\sigma^{z}\psi_{1}+\frac{1}{\alpha}aEM\omega \sigma^{z} \psi_{1}-M^{2}\omega ^{2}r^{2}\psi_{1}\notag \\
& +2M\omega \sigma^z\frac{1}{2\alpha}\left( 1-\alpha \right) \sigma^z\psi _{1}=0.\label{eav}
\end{align}
Equation (\ref{eav}) describes the motion of the Dirac oscillator in the presence of a uniform magnetic field and the Aharonov-Bohm potential in metric space-time (\ref{metric}), i.e., under topological and noninertial effects. Keeping in mind that $\psi_{1}$ is a bispinor and taking into account the structure of the Pauli matrices in cylindrical coordinates, we adopt the solutions with the form
\begin{equation}
\psi _{1}\left( r,\varphi \right)=\left(\begin{array}{c}
\psi _{a}\left( r,\varphi \right)  \\
\psi _{b}\left( r,\varphi \right)
\end{array}
\right) =\left(
\begin{array}{c}
e^{im\varphi }f\left( r\right)  \\
ie^{i\left( m+1\right) \varphi}g\left( r\right)
\end{array}
\right),\label{P1}
\end{equation}
where $\psi_{a}$ refers to the spin up component,  while $\psi_{b}$ is related to the down spin component. 	
Since we are only interested in regular solutions at the origin, we neglect the quantity $\frac{1}{\alpha r}\sigma ^{z}e\left[\partial_{r}\left(A_{\varphi ,2}\right) \right] \psi_{1}$ in Eq. (\ref{eav}). Thus, substituting (\ref{P1}) in Eq. (\ref{eav}) together with Eqs. (\ref{potentialv}) and (\ref{fields}) and reorganizing the terms, we obtain the following second order differential equation for $f(r)$:
\begin{equation}
\left( \frac{d^{2}}{dr^{2}}+\frac{1}{r}\frac{d}{dr}-\frac{L_{+}^{2}}{\alpha^{2}r^{2}}-M^{2}\Omega_{+} ^{2}r^{2}+\varepsilon_{+} \right) f\left( r\right) =0,
\label{rd}
\end{equation}%
where
\begin{equation}
L_{+}=\ell_{+} +aE, \;\text{with}\;\; 
\ell_{+} =m-\phi +\frac{1}{2}\left( 1-\alpha \right), \label{maef}
\end{equation}
is the effective angular momentum,
\begin{equation}
\Omega_+ =\omega +\frac{\omega _{c}}{2},\label{Omeff}
\end{equation}
is the effective frequency, with $\omega_{c}=eB/M$ defining the cyclotron frequency, and
\begin{equation}
\varepsilon_{+} =E^{2}-M^{2}+2M\Omega_{+} \left( 1+\frac{L_{+}}{\alpha}\right).
\end{equation}
It can be shown through a simple change of variables that Eq. (\ref{rd}) can be written in the form of a confluent hypergeometric differential equation, whose solution is well known. Thus, the non-normalized solution for $\psi _{a}\left( r,\varphi \right)$ is 
\begin{align}
&\psi _{a}\left( r,\varphi \right) =C_{nm}\left( M\Omega_{+} \right) ^{\frac{1}{2}+{\frac{\left\vert L_{+}\right\vert }{2\alpha}}}{r}^{{\frac{\left\vert	L_{+}\right\vert }{\alpha }}}e{^{-\frac{1}{2}\,M\Omega_{+} {r}^{2}}}e^{im\varphi }\notag \\
& \times {_{1}F_{1}}\left(\left(\frac{1}{2}+{\frac{\left\vert L_{+}\right\vert}{2\alpha }}\right)-{\frac{\varepsilon}{4M\Omega_{+} }},\,1+{\frac{\left\vert L_{+}\right\vert }{\alpha }},\,M\Omega_{+} {r}^{2}\right),\label{pphi}
\end{align}
where ${_{1}F_{1}}\left(a,b,z \right)$ denotes the confluent hypergeometric function of the first kind or Kummer's function $M(a,b,z)$ and $C_{nm}$ is the normalization constant. The hypergeometric function ${_{1}F_{1}}\left(a,b,z \right)$ has various important properties to assist us in our description. One of them emerges from studying the asymptotic behavior of the solution (\ref{pphi}) when we search for bound state solutions, which reveals a divergent behavior for large values of its argument, namely
\begin{equation}
{{_{1}F_{1}}\left( a,\,b,\,z\right) \approx }\frac{\Gamma \left( b\right) }{\Gamma \left( a\right) }e^{z}z^{a-b}\left[1+O\left( \left\vert z\right\vert
^{-1}\right) \right].
\end{equation}
Because of this divergent behavior of the function ${_{1}F_{1}}\left(a,b,z \right)$ for large values of its argument, bound states solutions can only be obtained by imposing that this function becomes a polynomial of degree $n$, where $n\in\mathbb{Z}^{*}$, with $\mathbb{Z}^{*}$ denoting the set of the nonnegative integers. This is established by requiring that
\begin{equation}
\frac{1}{2}\left( 1+{\frac{\left\vert L_{+}\right\vert }{\alpha }}\right) -
\frac{\varepsilon_{+}}{4M\Omega_{+}}=-n,\label{cond}
\end{equation}
which allows us to write the bound state solutions in the form
\begin{align}
&\psi_{a}\left( r,\varphi \right)=C_{nm}\left( M\Omega_{+} \right)^{\frac{1}{2}+{\frac{\left\vert L_{+}\right\vert}{2\alpha}}}{r}^{{\frac{\left\vert L_{+}\right\vert }{\alpha}}}e{^{-\frac{1}{2}\,M \Omega_{+} {r}^{2}}}e^{im\varphi }
\notag \\
& \times {_{1}F_{1}}\left(-n,\,1+{\frac{\left\vert L_{+}\right\vert }{\alpha}},\,M\Omega_{+} {r}^{2}\right). \label{sbs}
\end{align}
Analyzing Eq. (\ref{cond}), we see it depends on energy through the absolute value of the effective angular moment $L_{+}$. Then, to extract the expression for the energy eigenvalues to the oscillator, we must solve the condition (\ref{cond}) for $E$ considering $L_{+}>0$ and $L_{+}<0$, respectively. We obtain
\begin{align}
&E_{n,+}^{\left( >\right) }=\pm \sqrt{4nM\Omega_{+} +M^{2}} \hspace{0.3cm} \mbox{for} \hspace{0.2cm} L_+>0, \label{cp1}\\
&E_{nm,+}^{\left( <\right) }=-\frac{2M\Omega_{+} a}{\alpha } \notag \\&\pm \frac{1}{\alpha }%
\sqrt{4a^{2}M^{2}\Omega_{+}^{2}+4M\Omega_{+} \alpha \left( n\alpha -\ell_+ \right) +\alpha^{2}M^{2}} \label{cp2}
\end{align}
for $L_+<0$.
\begin{figure}[!b]
\centering	
\includegraphics[scale=0.28]{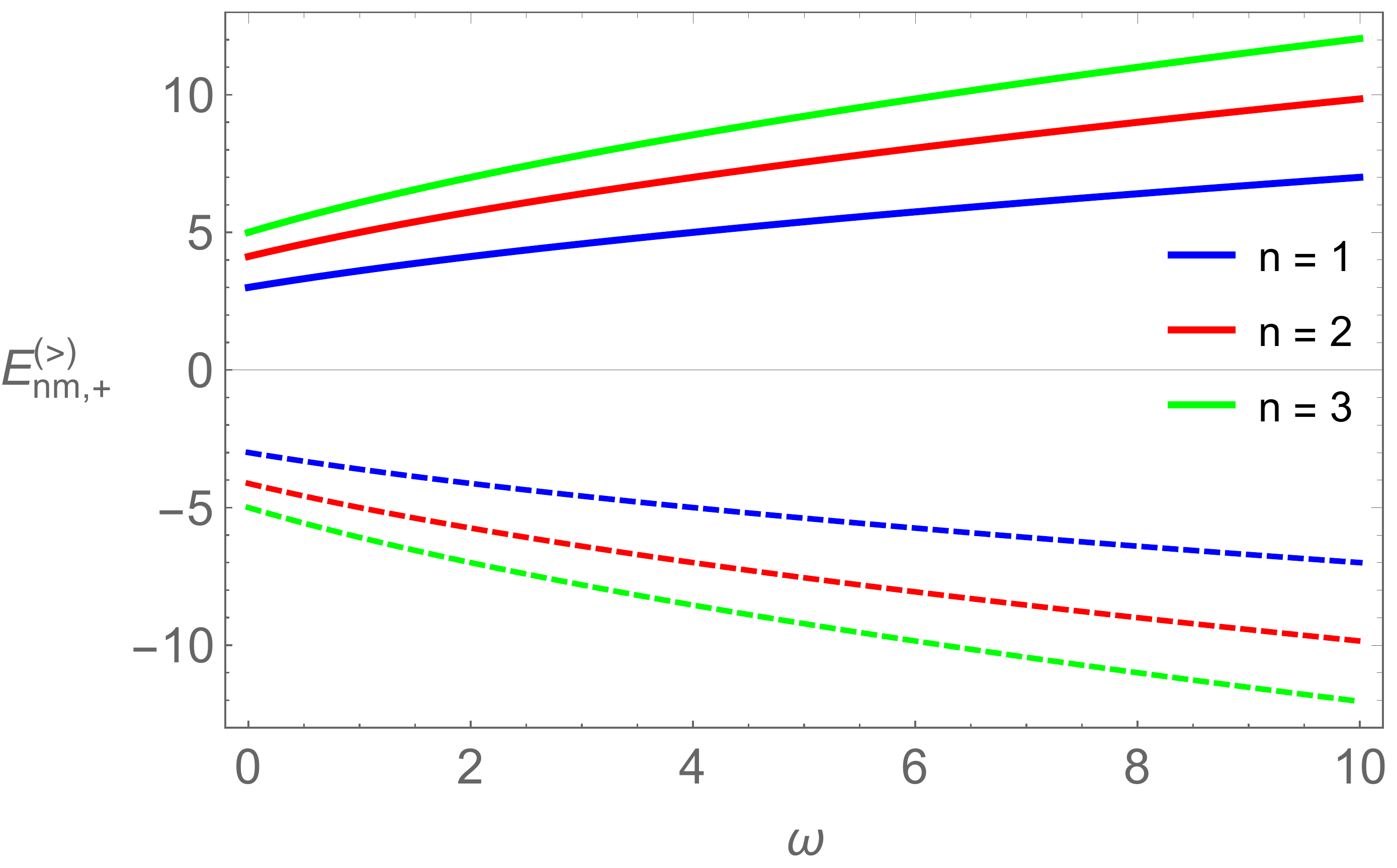}
\caption{Energy levels $E_{nm,+}^{(>)}$ (Eq. (\ref{cp1})) as a function of $\omega$.}
\label{Fig2D_P_ExFosci_Wc0}
\end{figure}
\begin{figure}[!t]
\centering
\includegraphics[scale=0.18]{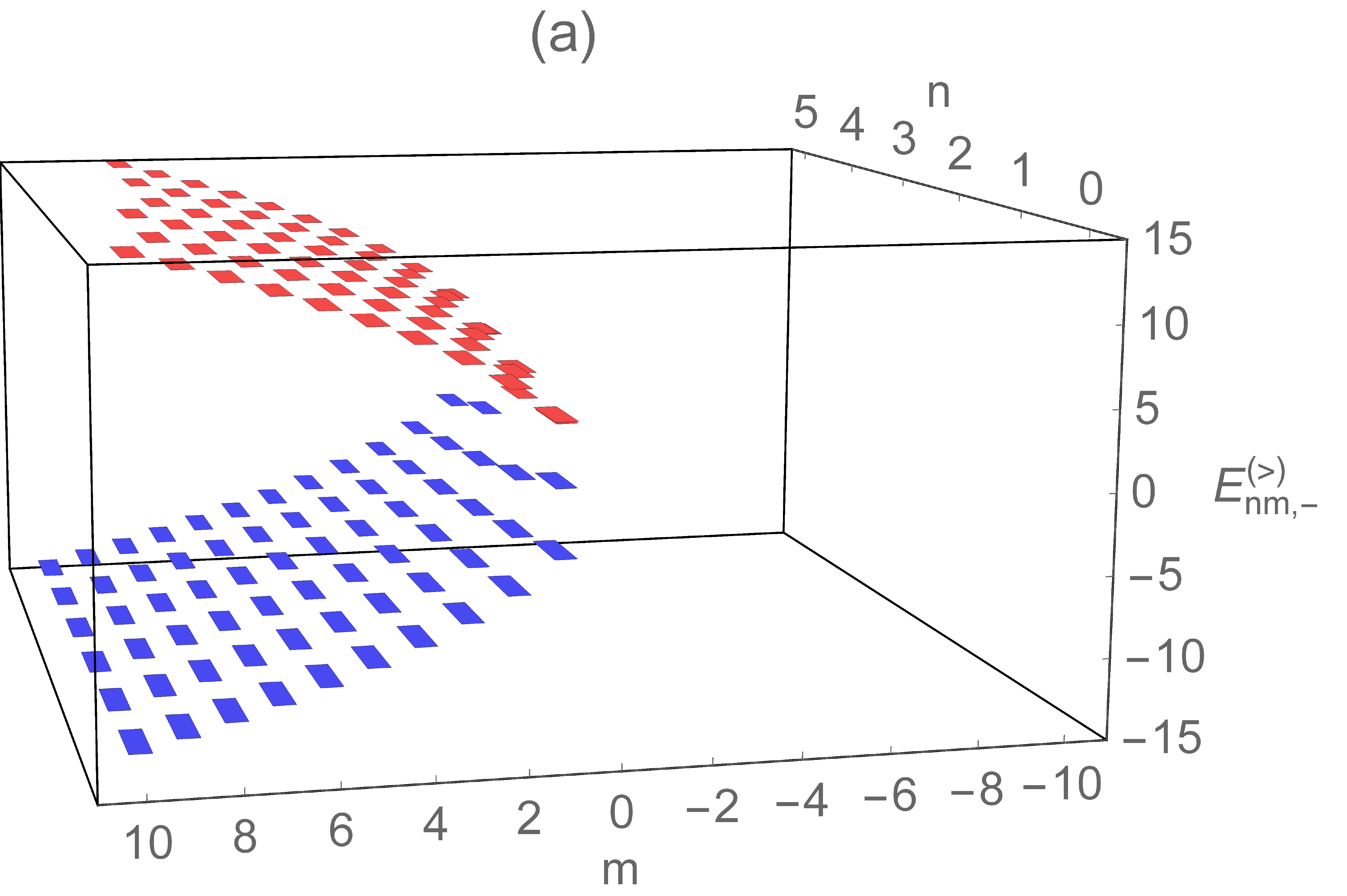}\vspace{0.3cm}
\includegraphics[scale=0.18]{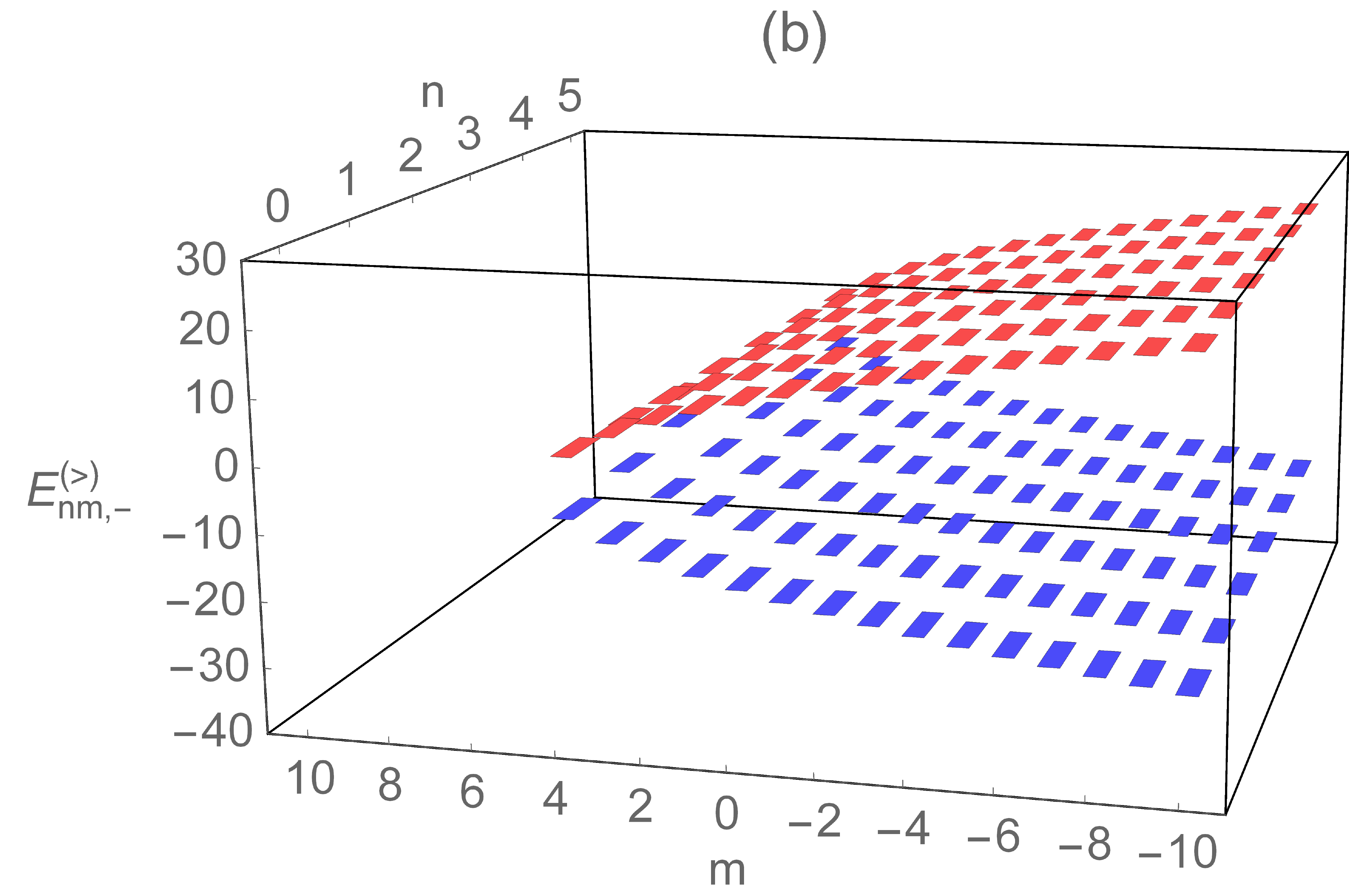}
\caption{(Color online)
Profile of $|E_{nm,-}^{(>)}|$ (Eq. (\ref{cpm1_2})) as a function of $n$ and $m$. In panel (a), $a=0.1$, $\alpha=0.4$, $\phi=1$, $\omega=3$ and $\omega_{c}=2$ while in panel (b), $a=0.1$, $\alpha=0.2$, $\phi=5$, $\omega=3$ and $\omega_{c}=9.9$.}
\label{Fig3D_Enm}
\end{figure}
\begin{figure}[!h]
\centering	
\includegraphics[scale=0.28]{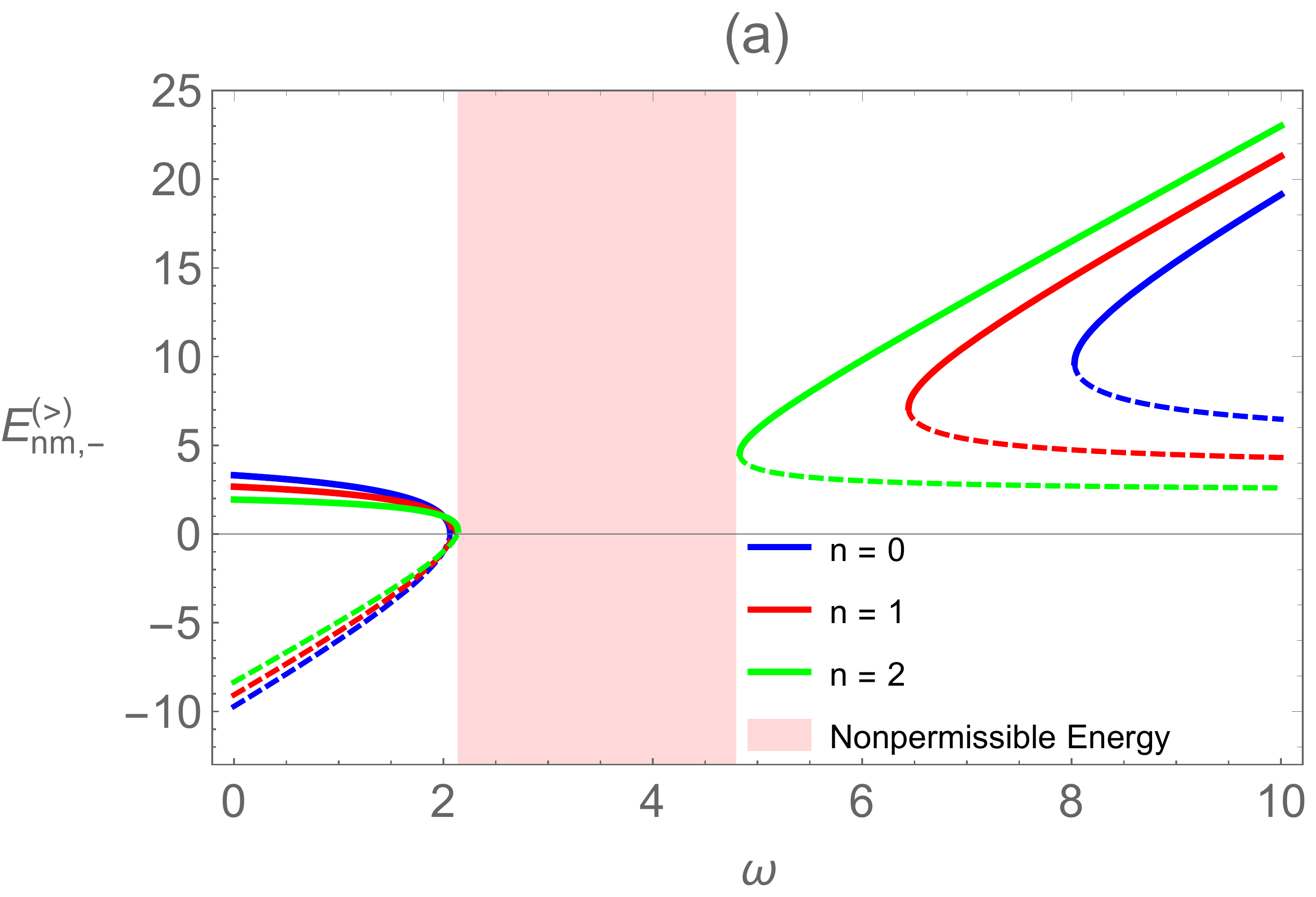}\vspace{0.2cm}
\includegraphics[scale=0.28]{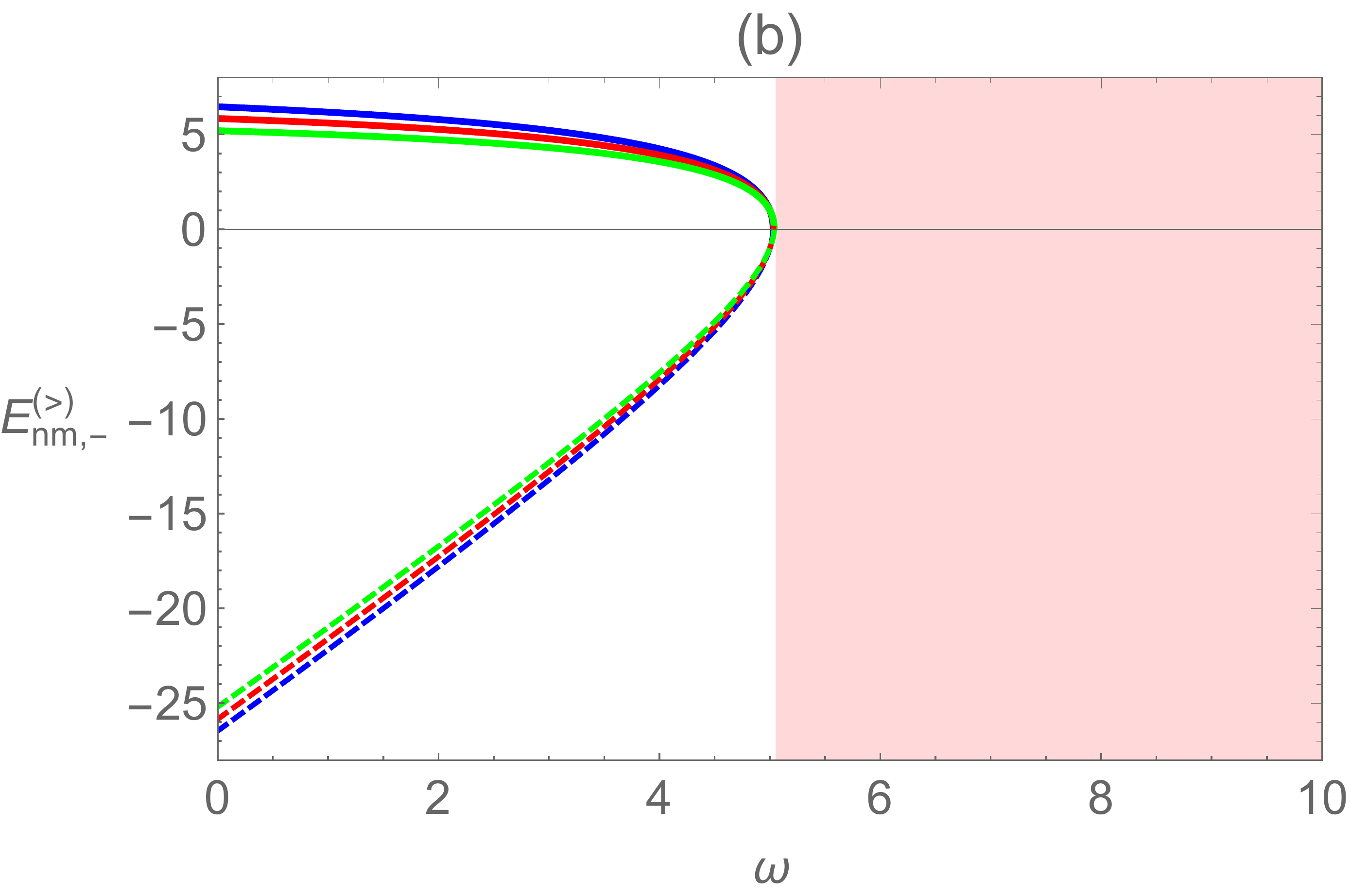}\vspace{0.2cm}
\includegraphics[scale=0.28]{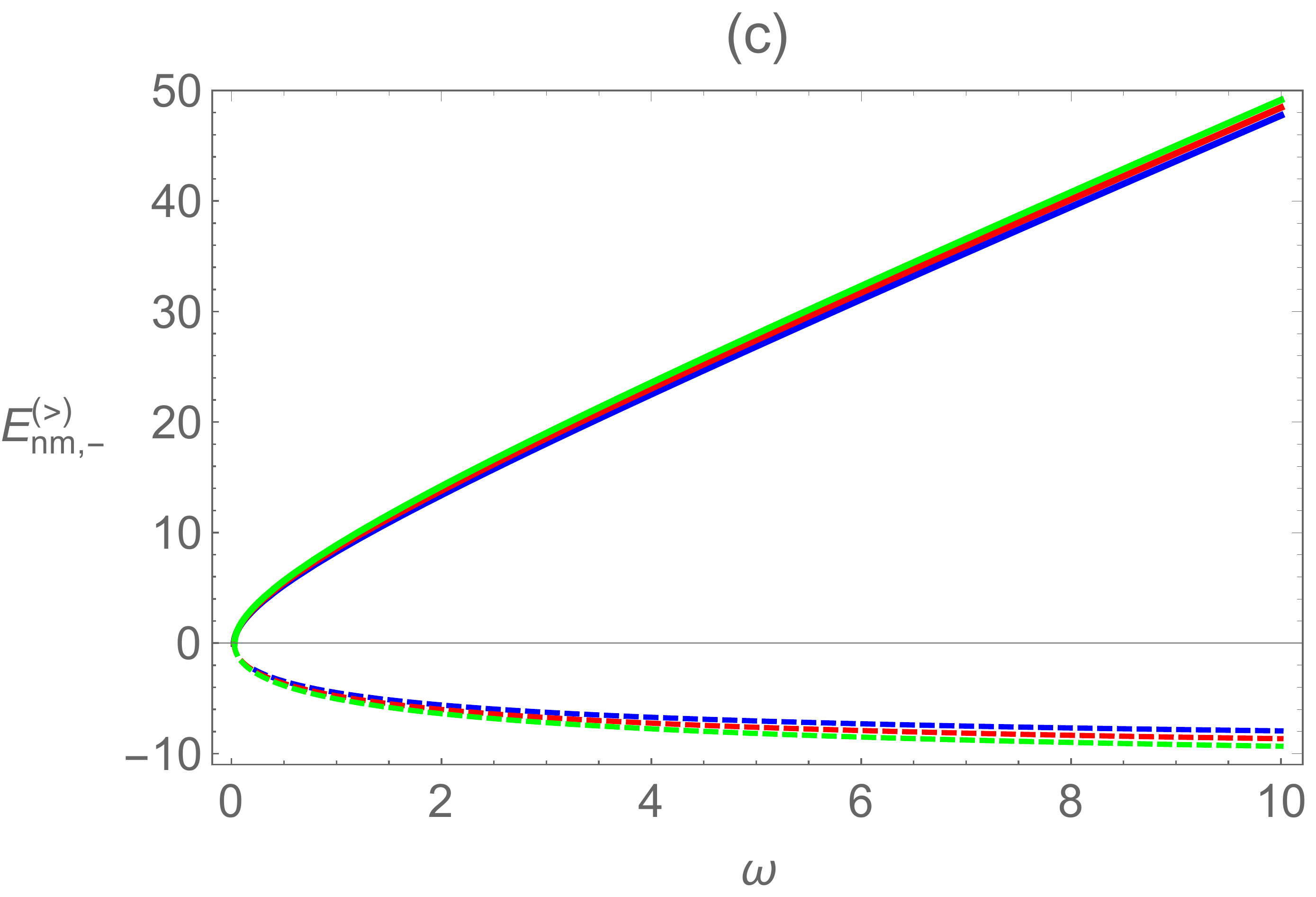}
\caption{Sketch of the energy levels $|E_{nm,-}^{(>)}|$ (Eq. (\ref{cpm1_2})) as a function of $\omega$. In panel (a), $m=-2$, $a=0.4$, $\alpha=0.5$, $\phi=0.7$ and $\omega_{c}=4$. In panel (b), $m=5$, $a=0.5$, $\alpha=0.5$, $\phi=10$ and $\omega_{c}=10$. In panel (c), $m=5$, $a=0.5$, $\alpha=0.5$, $\phi=1$ and $\omega_{c}=0.1$.}
\label{Fig2DP_E_W_osc}
\end{figure}
\begin{figure}[!h]
\centering	
\includegraphics[scale=0.28]{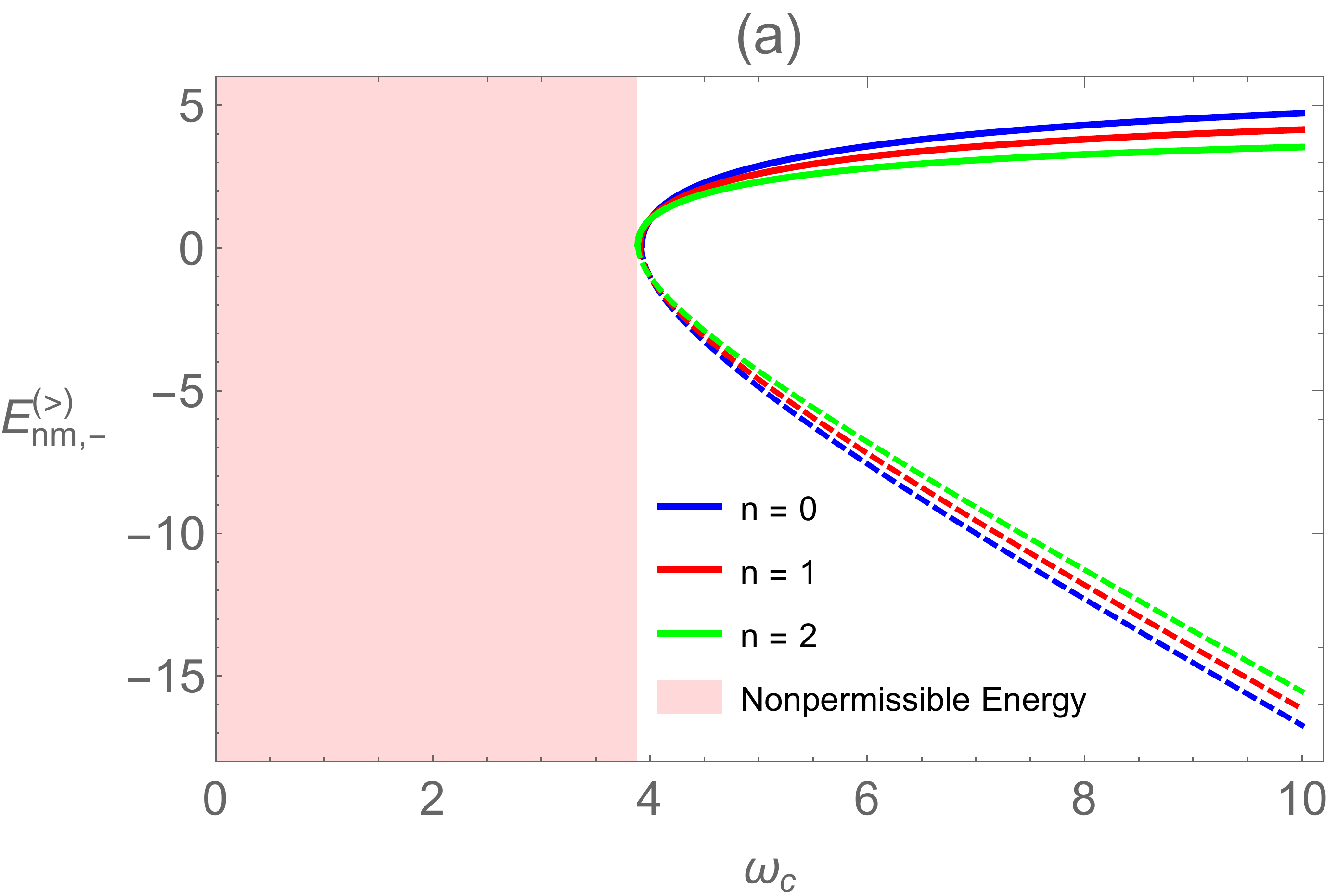}\vspace{0.2cm}
\includegraphics[scale=0.28]{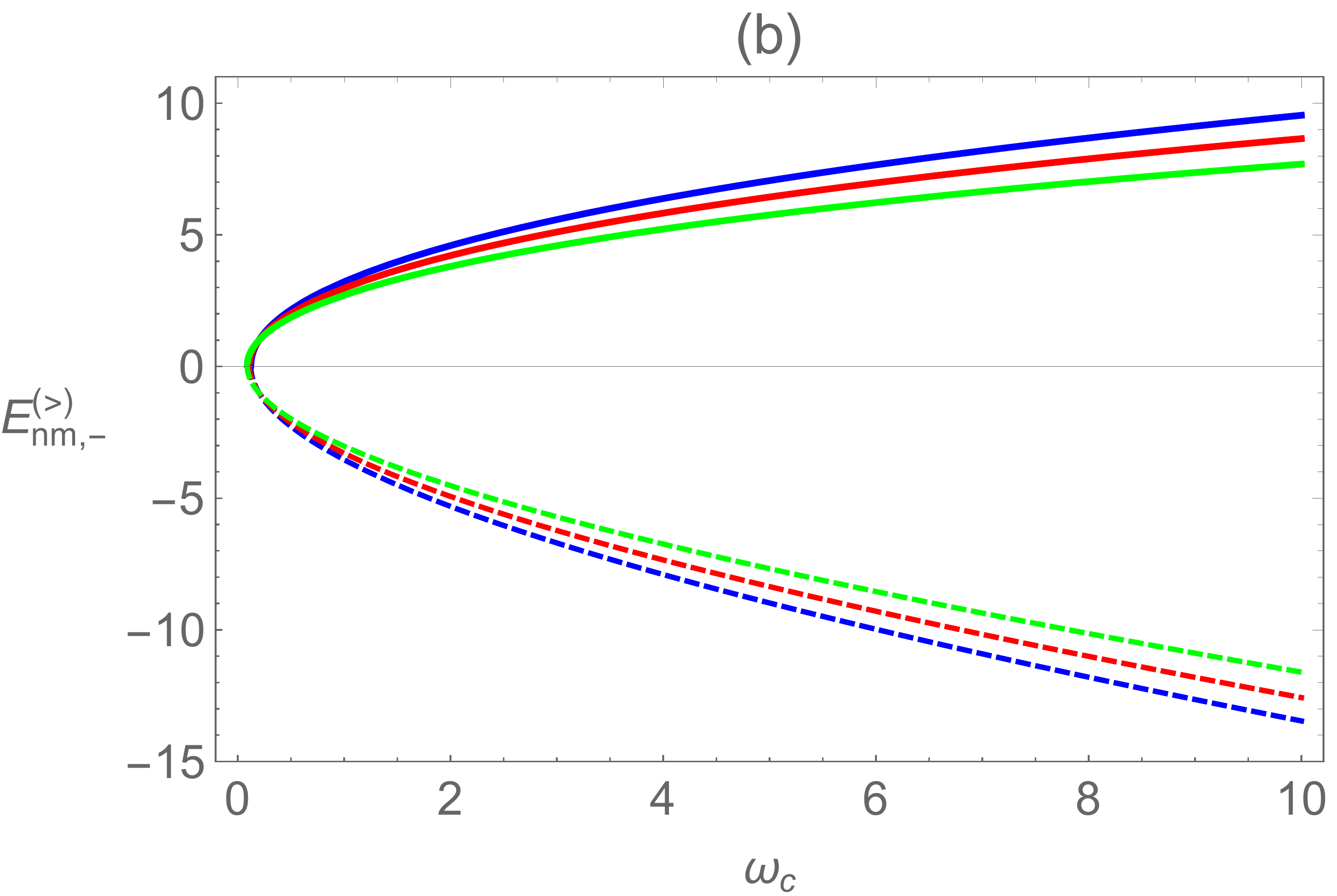}\vspace{0.2cm}
\includegraphics[scale=0.28]{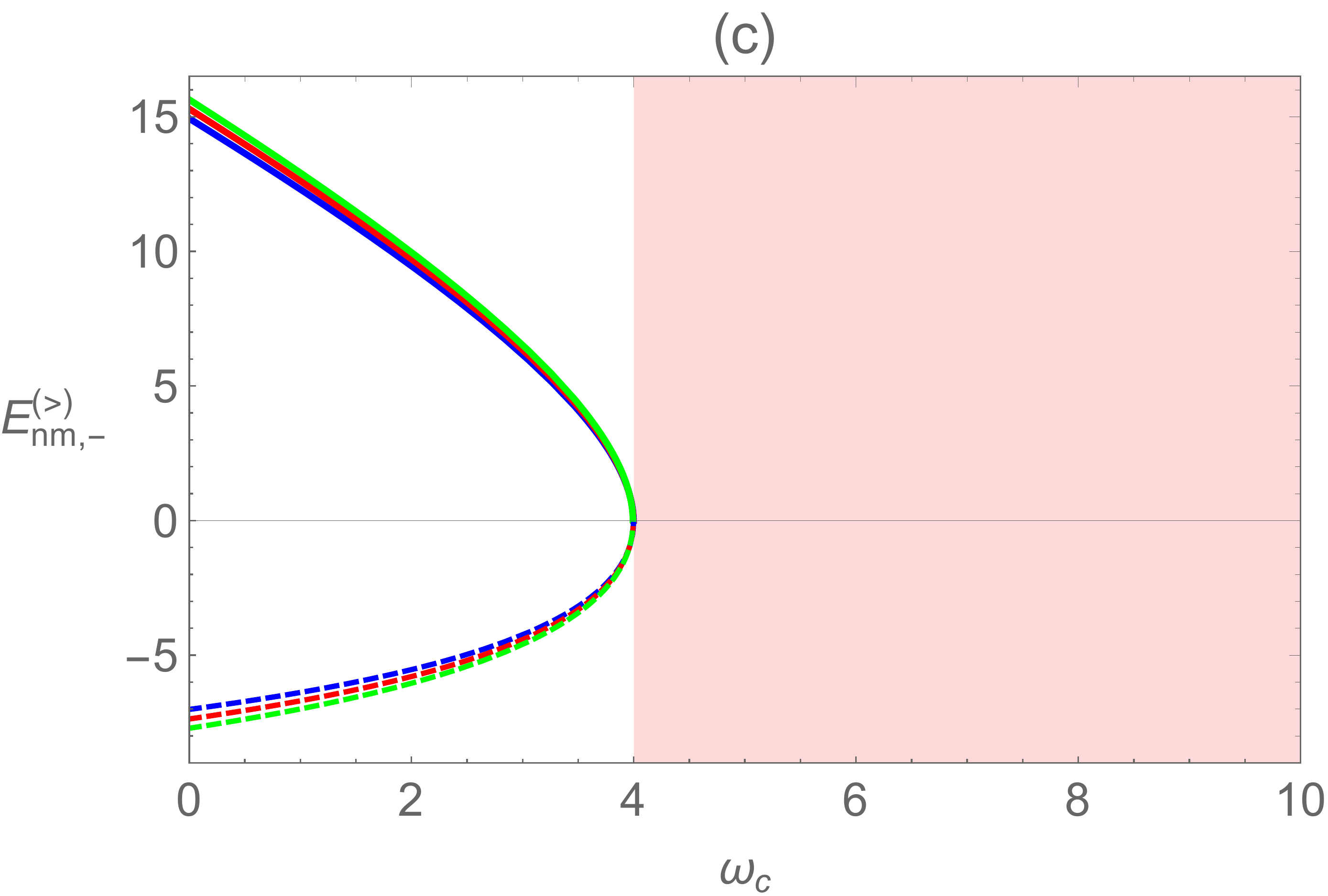}
\caption{Sketch of the energy levels $|E_{nm,-}^{(>)}|$ (Eq. (\ref{cpm1_2})) as a function of $\omega_{c}$. In panel (a), $m=-2$, $a=0.5$, $\alpha=0.5$, $\phi=2$ and $\omega=2$. In panel (b), $m=-2$, $a=0.1$, $\alpha=0.5$, $\phi=1$ and $\omega=0.1$. In panel (c), $m=6$, $a=0.5$, $\alpha=0.5$, $\phi=0.2$ and $\omega=1.98$.}
\label{Fig2DP_E_Wciclo}
\end{figure}

The equation for $g(r)$, corresponding to the down spin component, is similar to the Eq. (\ref{maef}). Now, we have
\begin{equation}
\left( \frac{d^{2}}{dr^{2}}+\frac{1}{r}\frac{d}{dr}-\frac{L_{-}^{2}}{\alpha^{2}r^{2}}-M^{2}\Omega_{-} ^{2}r^{2}+\varepsilon_{-} \right) g\left( r\right) =0,\label{rd_2}
\end{equation}%
where
\begin{equation}
L_{-}=\ell_{-} +aE, \;\text{with}\;\; \ell_{-} =\left( m+1 \right)-\phi -\frac{1}{2}\left( 1-\alpha \right), \label{maef_2}
\end{equation}
and	
\begin{equation}
\varepsilon_{-} =E^{2}-M^{2}+2M\Omega_{-} \left( 1-\frac{L_{-}}{\alpha}\right),
\end{equation}	
\begin{equation}
\Omega_{-} =\omega -\frac{\omega_{c}}{2}.\label{Omeff_2}
\end{equation}	
The corresponding energies are	
\begin{align}
&E_{nm,-}^{\left( >\right)} =\frac{2Ma\,\Omega_{-}}{\alpha}\notag \\&\pm \frac{1}{\alpha }
\sqrt{4a^{2}M^{2}\Omega_{-} ^{2}+4M\Omega_{-} \alpha \left( n\alpha +\ell_{-} \right) +\alpha^{2}M^{2}}, \label{cpm1_2}
\end{align}
for $L_{-}>0$, and
\begin{equation}
E_{n,-}^{\left(<\right)} =\pm \sqrt{4nM\Omega_{-} +M^{2}} \label{cpm2_2}
\end{equation}
for $L_{-}<0.$ 
Energies (\ref{cp1}), (\ref{cp2}), (\ref{cpm1_2}) and (\ref{cpm2_2}) constitute the energy spectrum of the Dirac oscillator for the model in question, and is a generalization of that addressed in Ref. \cite{EPJC.2019.79.311} for the case where we have the presence of external magnetic fields. We can study in detail both the positive and negative energies of the particle. A first analysis reveals that they depend on all the physical parameters involved in the problem. Depending on the choice we make for the parameters, these energies provide exotic and interesting results. This is related to the fact that even a small change in any of these parameters can imply an abrupt change in the profile of the energy spectrum. To understand how rotation, Aharonov-Bohm flow, uniform field, and curvature affect the oscillator's energy levels, we opt to study some particular configurations, which are defined from the choice for the set of parameters.
In principle, we can analyze these energies seeking to access their most relevant properties, such as the profile for strong and weak fields, high and low rotations, etc.  
 In the discussions below we use $M=1$. We can clearly see that the energies (\ref{cp1}) and (\ref{cpm2_2}) are symmetrical for any $n$, $\Omega_{+}$ and $\Omega_{-}$. In the case of Eq. (\ref{cp1}), $|E_{nm,+}^{(>)}|$ increases when $n$ increases (Fig. \ref{Fig2D_P_ExFosci_Wc0}). If we make $\omega_{c}=0$ in Eqs. (\ref{cp1}) and (\ref{cpm2_2}), we recover the result (48) from Ref. \cite{EPJC.2019.79.311}. Here, the most interesting characteristics are manifested in the energies (\ref{cp2}) and (\ref{cpm1_2}). The analysis of these energies reveals they have similar spectrums. This way, we prefer to examine energy (\ref{cpm1_2}). It can be shown that this energy is conditioned to $4a^{2}M^{2}\Omega_{-} ^{2}+4M\Omega_{-} \alpha \left( n\alpha +\ell_{-} \right) +\alpha^{2}M^{2} \geqslant 0$. This condition combined with the quantity $(2M\Omega_{-} a)/\alpha$ is the key to understanding the results. One of the most fascinating manifestations here is the existence of non-permissible energy states. For example, for $a=0.1$, $\alpha=0.4$, $\phi=1$, $\omega=3$ and $\omega_{c}=2$, the energies as a function of $m$ and $n$ show that states with $m<0$ are forbidden (Fig. \ref{Fig3D_Enm}(a)) while for parameters $a=0.1$, $\alpha=0.2$, $\phi=5$, $\omega=3$ and $\omega_{c}=9.9$ the energies are allowed for any $m$ (Fig. \ref{Fig3D_Enm}(b)). 
\begin{figure}[!t]
\centering	
\includegraphics[scale=0.28]{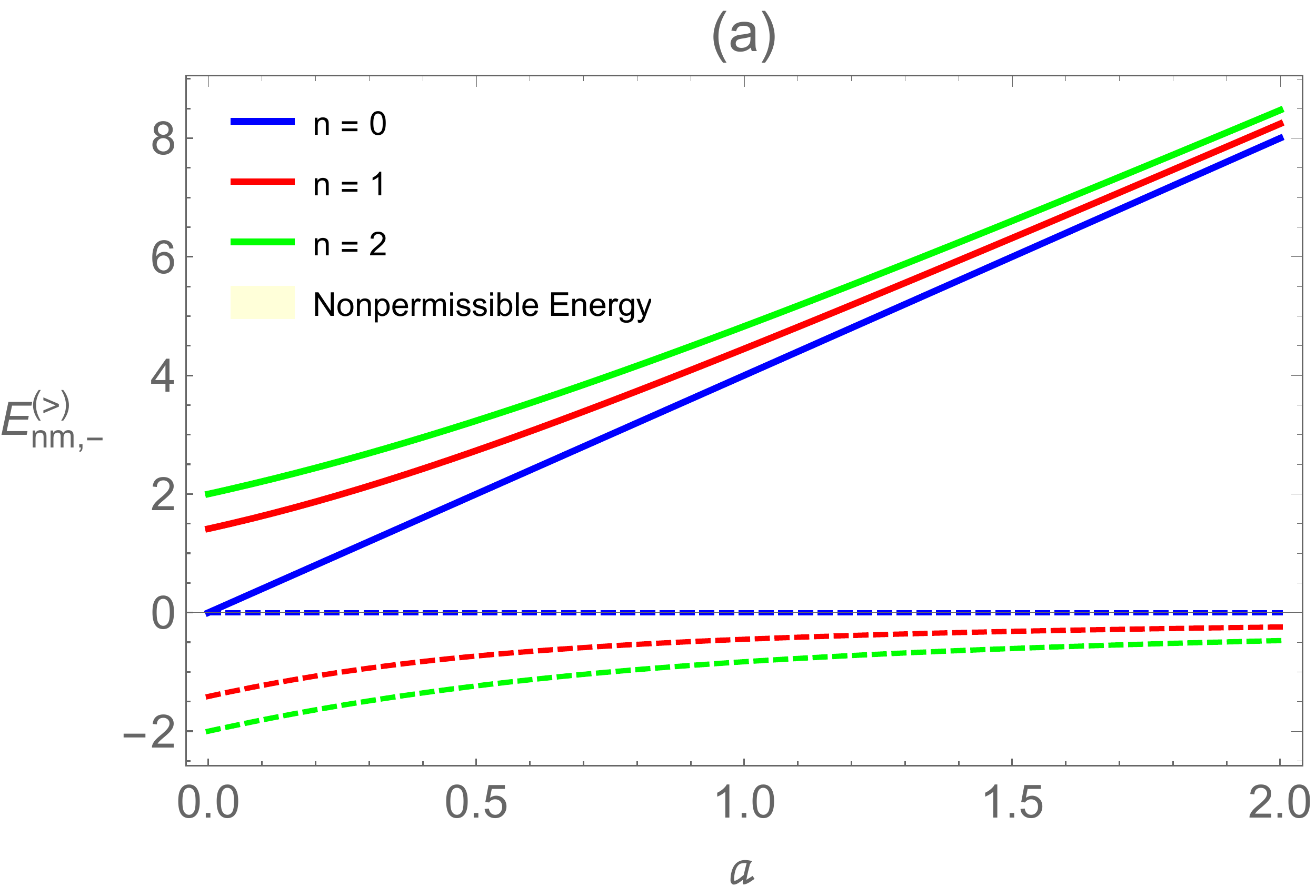}\vspace{0.2cm}
\includegraphics[scale=0.28]{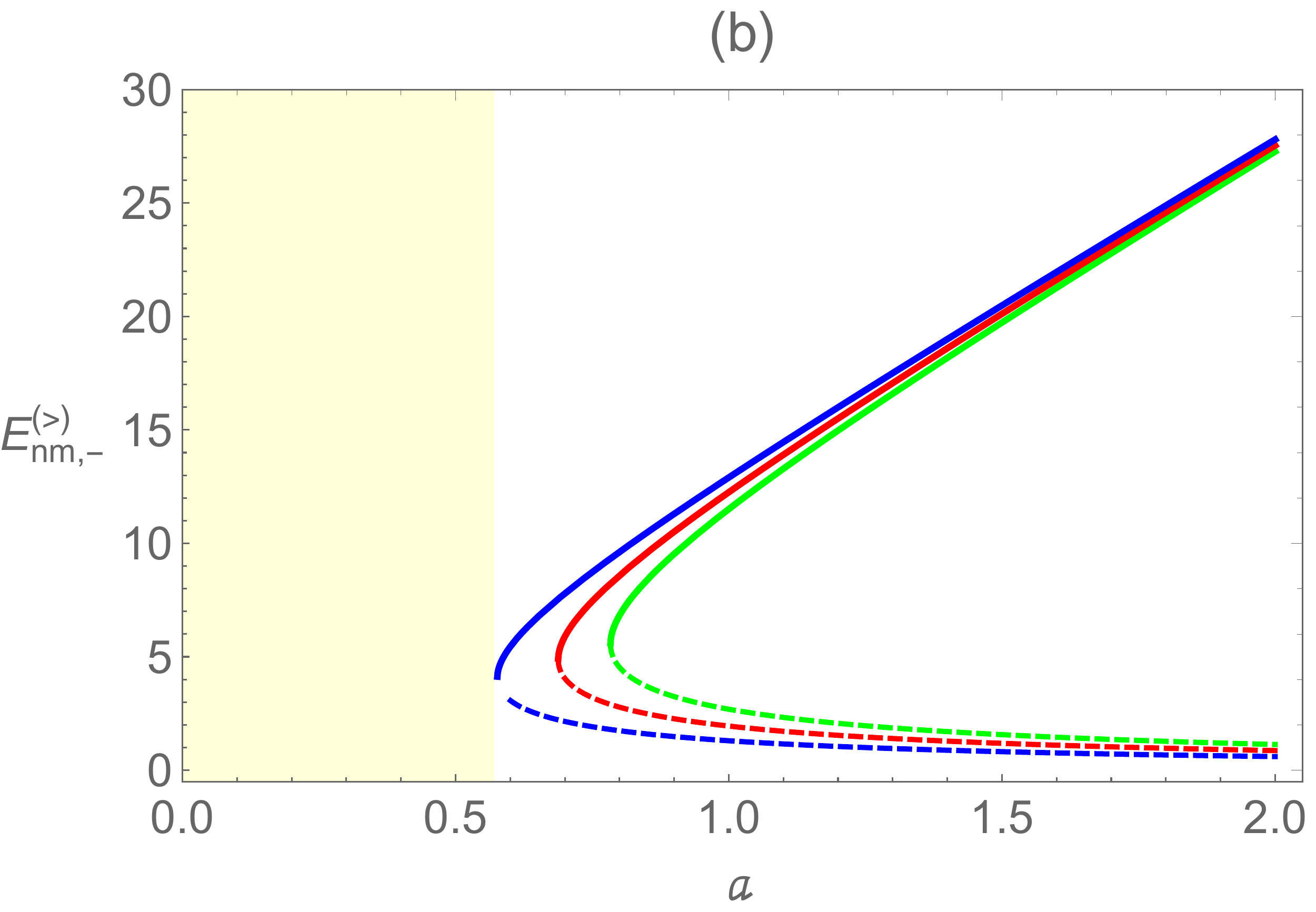}\vspace{0.2cm}
\includegraphics[scale=0.29]{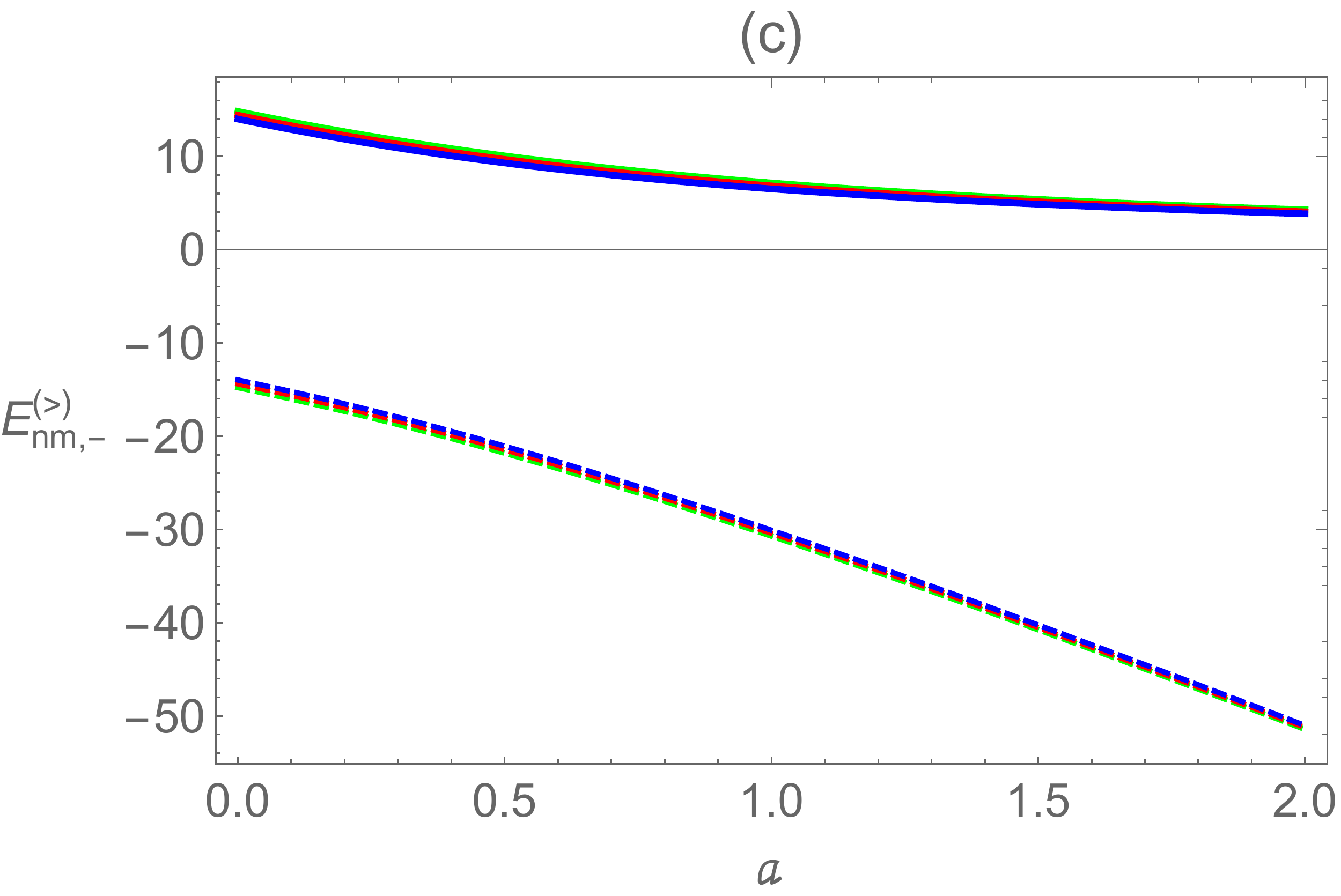}
\caption{Sketch of the energy levels $|E_{nm,-}^{(>)}|$ (Eq. (\ref{cpm1_2})) as a function of $a$. In panel (a), $m=2$, $\alpha=0.5$,  $\phi=3$, $\omega=2$ and $\omega_{c}=3$. In panel (b), $m=-2$, $\alpha=0.5$,  $\phi=1$, $\omega=2$ and $\omega_{c}=0.45$. In panel (c), $m=-1$, $\alpha=0.5$,  $\phi=9$, $\omega=2$ and $\omega_{c}=9.9$.}
\label{Fig2DP_Erot}
\end{figure}
\begin{figure}[!ht]
\centering	
\includegraphics[scale=0.28]{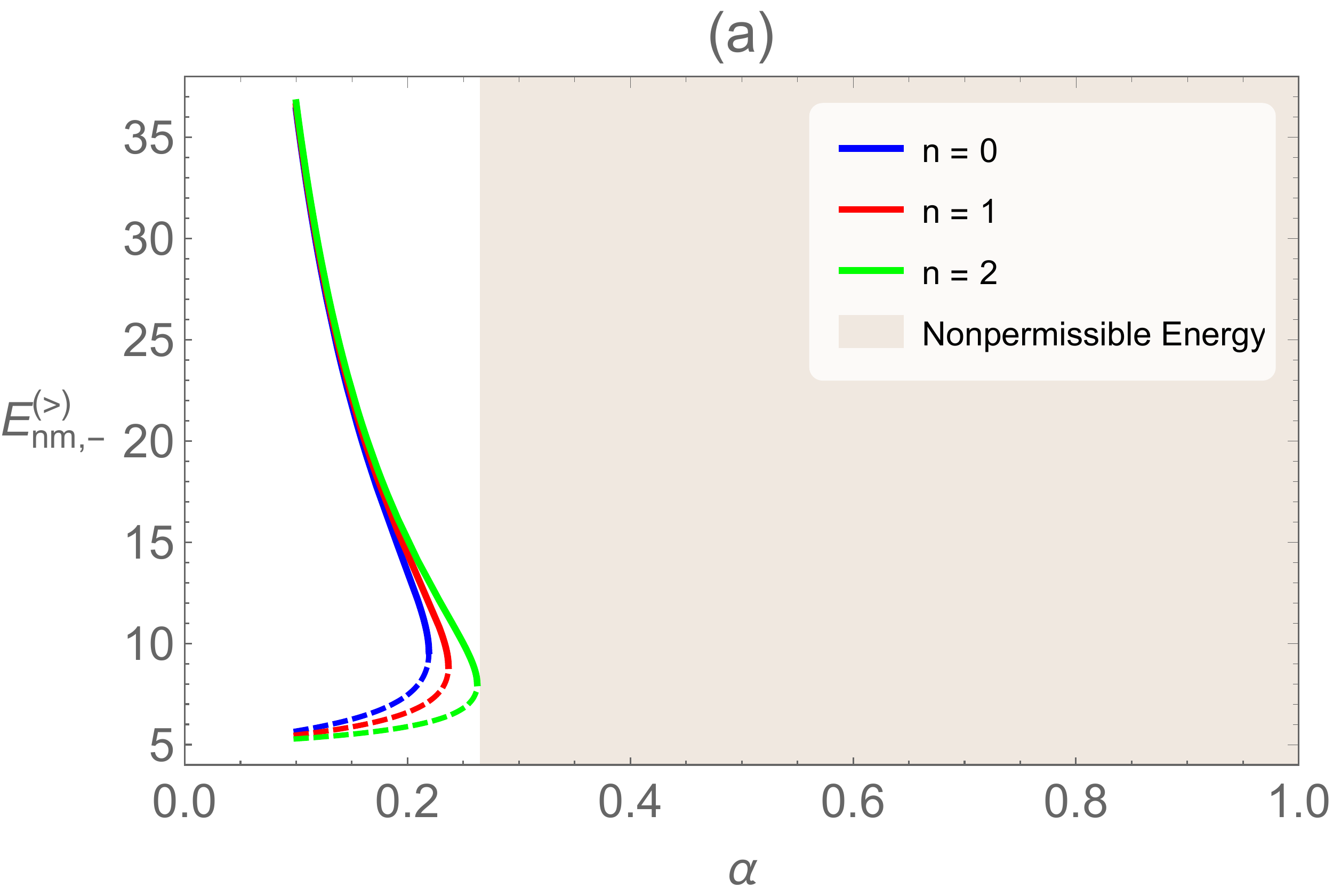}\vspace{0.2cm}
\includegraphics[scale=0.28]{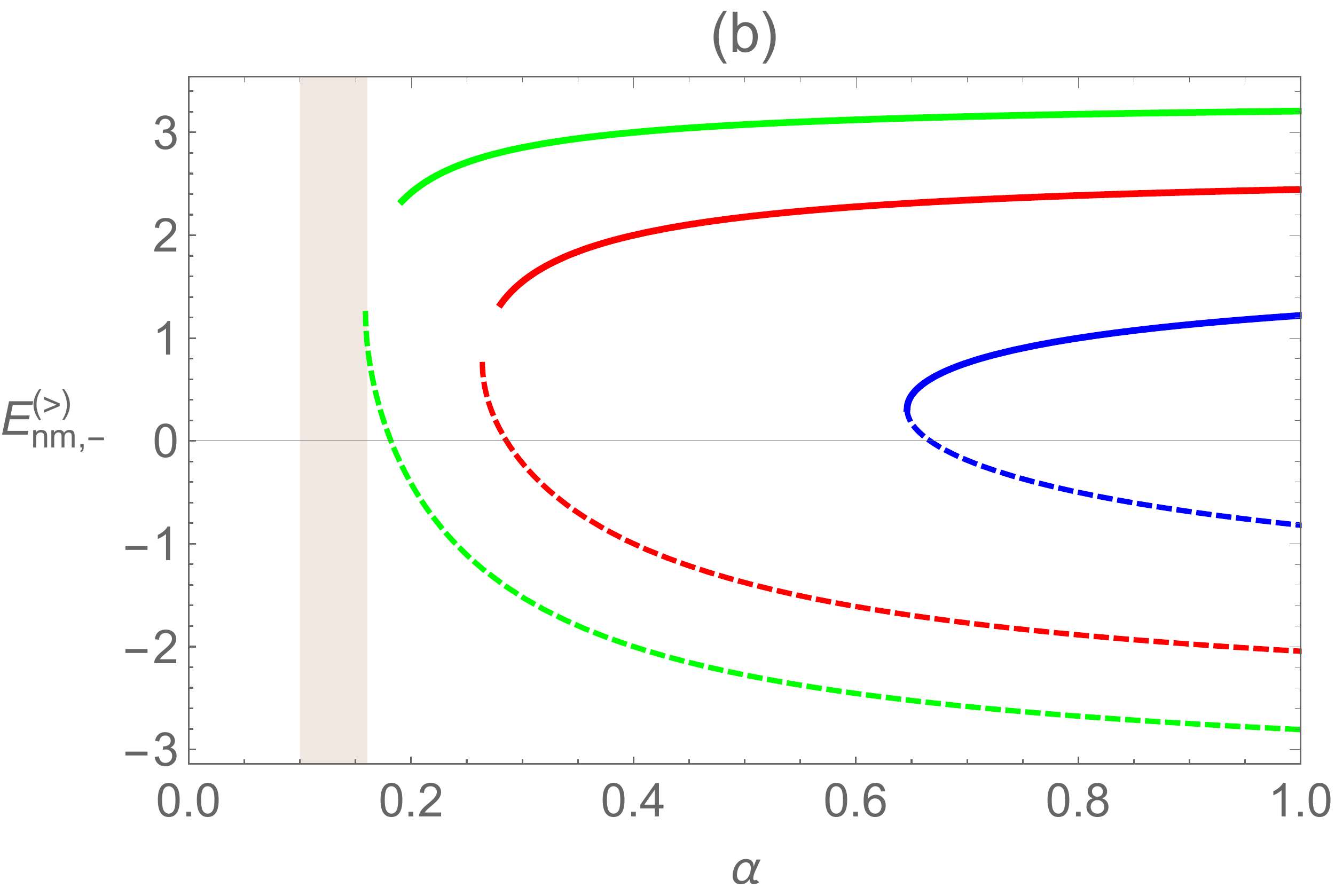}\vspace{0.2cm}
\includegraphics[scale=0.29]{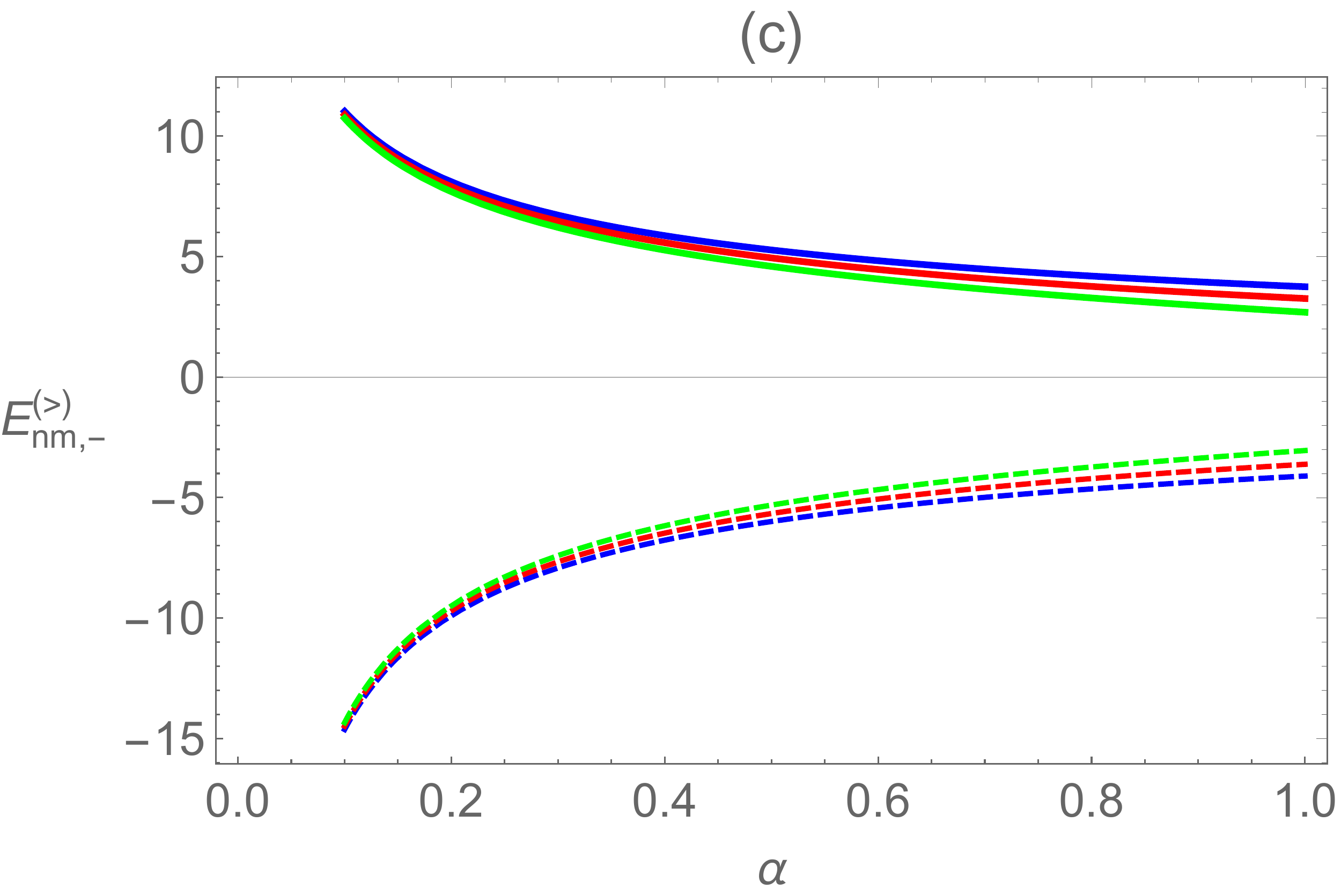}
\caption{Sketch of the energy levels  $|E_{nm,-}^{(>)}|$ (Eq. (\ref{cpm1_2})) as a function of $\alpha$. In panel (a), $m=2$, $a=0.7$, $\phi=6$, $\omega=2$ and $\omega_{c}=1$. In panel (b), $m=1$, $a=0.1$, $\phi=2$, $\omega=2$ and $\omega_{c}=2$. In panel (c), $m=1$, $a=0.1$, $\phi=6$, $\omega=4$ and $\omega_{c}=9.8$.}
\label{Fig2DP_Ealpha}
\end{figure}
When we examine the profile of (\ref{cpm1_2}) as a function of $\omega$ and $\omega_{c}$, we note other impressive characteristics. The sketch as a function of $\omega$ for $m=-2$, $a=0.4$, $\alpha=0.5$ and $\phi=0.7$ reveals the existence of forbidden energy states for some values of $\omega$ (Fig. \ref{Fig2DP_E_W_osc}(a)). In the range $2.0<\omega<4.8$, no energy state is permissible. One can show that for $m=-4$ and $m=-5$ (maintaining the other parameters) this interval increases until the spectrum completely assumes the profile of the region with $\omega<2$. In this region, there is an inversion between the energy levels, i.e., the higher energy states are those with $n=0$, $n=1$, $n=2$, respectively. In other words, $|E_{nm,-}^{(>)}|$ decreases when $n$ increases. This effect, however, is being undone when $\omega$ approaches $2$. For $\omega>4.8$, the opposite effect occurs, i.e., $|E_{nm,-}^{(>)}|$ increases when $n$ increases. Note that the forbidden energy range increases from the most energetic to the least energetic states. When both $\phi$ and $\omega_{c}$ are increased and $m=5$, the interval of $\omega$ giving nonpermissible energies is increased (Fig. \ref{Fig2DP_E_W_osc}(b)). For weak magnetic field and $\phi=1$, $|E_{nm,-}^{(>)}|$ increases when $n$ increases (Fig. \ref{Fig2DP_E_W_osc}(c)). As we can see in Fig. \ref{Fig2DP_E_W_osc}, the energy is not symmetrical like in Fig. \ref{Fig3D_Enm}. This characteristic is a physical implication due to noninertial effects on the electron motion. At least for the parameter values that we consider, the energy profile as a function of $\omega_{c}$ (Fig. \ref{Fig2DP_E_Wciclo}) also displays some effects equivalent to the profile of Fig. \ref{Fig2DP_E_W_osc}. For $m=-2$, $a=0.5$, $\alpha=0.5$, $\omega=2$ and $\phi=2$, $|E_{nm,-}^{(>)}|$ increases when $n$ is increased (Fig. \ref{Fig2DP_E_Wciclo}(a)). For this configuration, the range of physically acceptable energies starts from $\omega=3.8$, corresponding to the energy state with $n=2$. Next to this value, an inversion between energy states occurs rapidly when $\omega_{c}$ is increased, which leads to a new energy configuration, where the states with higher energy are those starting with $n=0$, etc. When we access the energy state with $\omega=0.1$, $a=0.1$, $\phi=1$ and maintaining the other parameters, the range of allowed energies includes all values of $\omega_{c}$ (Fig. \ref{Fig2DP_E_Wciclo}(b)). This allows us to affirm that for smaller values of $a$, both particle and anti-particle tend to access all energy states. On the other hand, this configuration changes when we consider $m=6$, $a=0.5$, $\alpha=0.5$, $\omega=1.98$ and $\phi=0.2$, revealing the profile of fig. \ref{Fig2DP_E_Wciclo}(c). In this case, $|E_{nm,-}^{(>)}|$ increases when $n$ is increased. However, it decreases when $w_{c}$ is increased. 

When we investigate the behavior of the energy as a function of $a$, forbidden energy intervals are also observed, but with other curious features (Fig. \ref{Fig2DP_Erot}). For example, for the choice $m=2$, $\alpha=0.5$, $\omega=2$ and $\phi=3$ and $\omega_{c}=3$, except the state with $n=0$, we see that $|E_{nm,-}^{(>)}|$ increases when $n$ increases (Fig. \ref{Fig2DP_Erot}(a)). This profile changes when we adopt the parameters $m=-2$, $\alpha=0.5$, $\omega=2$ and $\phi=1$ and $\omega_{c}=0.45$, where energies are now  allowed only for $a>0.55$ (Fig. \ref{Fig2DP_Erot}(b)). The most significant impact on this profile is the change of state with $m=2$ to $m=-1$. The change from $\omega_{c}=3$ to $\omega_{c}=0.45$ has only the effect of increasing the nonpermissible energy range. A third configuration (making $m=-1$, $\phi=9$, $\omega_{c}=9.9$ and keeping the other parameters fixed) shows an indifference between the energy states, but we still observe the increase of $|E_{nm,-}^{(>)}|$ (Fig. \ref{Fig2DP_Erot}(c)).

Now, we discuss the behavior of (\ref{cpm1_2}) as a function of $\alpha$. By controlling $\alpha$, we can access localized energy states for an appropriate choice of the other parameters (Fig. \ref{Fig2DP_Ealpha}). In Fig. \ref{Fig2DP_Ealpha}(a), energy states with $m=2$, $a=0.7$, $\phi=6$, $\omega=2$ and $\omega_{c}=1$ are depicted. For the state with $n=2$, which is the largest energy state, a critical value of $\omega$ at $0.25$ occurs. Clearly, we see that significant values of $\alpha$ lead to localized energy states. If we change the configuration to $m=1$, $a=0.1$, $\phi=2$, $\omega=2$ and $\omega_{c}=2$, the insignificant curvature range is now located to the left of the spectrum, with a critical value occurring at $\alpha=0.15$, which delimits the state with $n=2$ (Fig. \ref{Fig2DP_Ealpha}(b)). For this configuration, $|E_{nm,-}^{(>)}|$ increases when $n$ and $\alpha$ are increased. We also observe that a inversion between states occurs when we keep the same $m$ and $a$ of Fig. \ref{Fig2DP_Ealpha}(b), and use $\phi=6$, $\omega=4$ and $\omega_{c}=9.8$ (Fig. \ref{Fig2DP_Ealpha}(c)). Similarly to the others cases we saw above, this configuration also shows that $|E_{nm,-}^{(>)}|$ decreases when $n$ and $\alpha$ are increased. In addition, we also see a reduction of the spacing between the energy levels.
\begin{figure}[!h]
\centering
\includegraphics[scale=0.19]{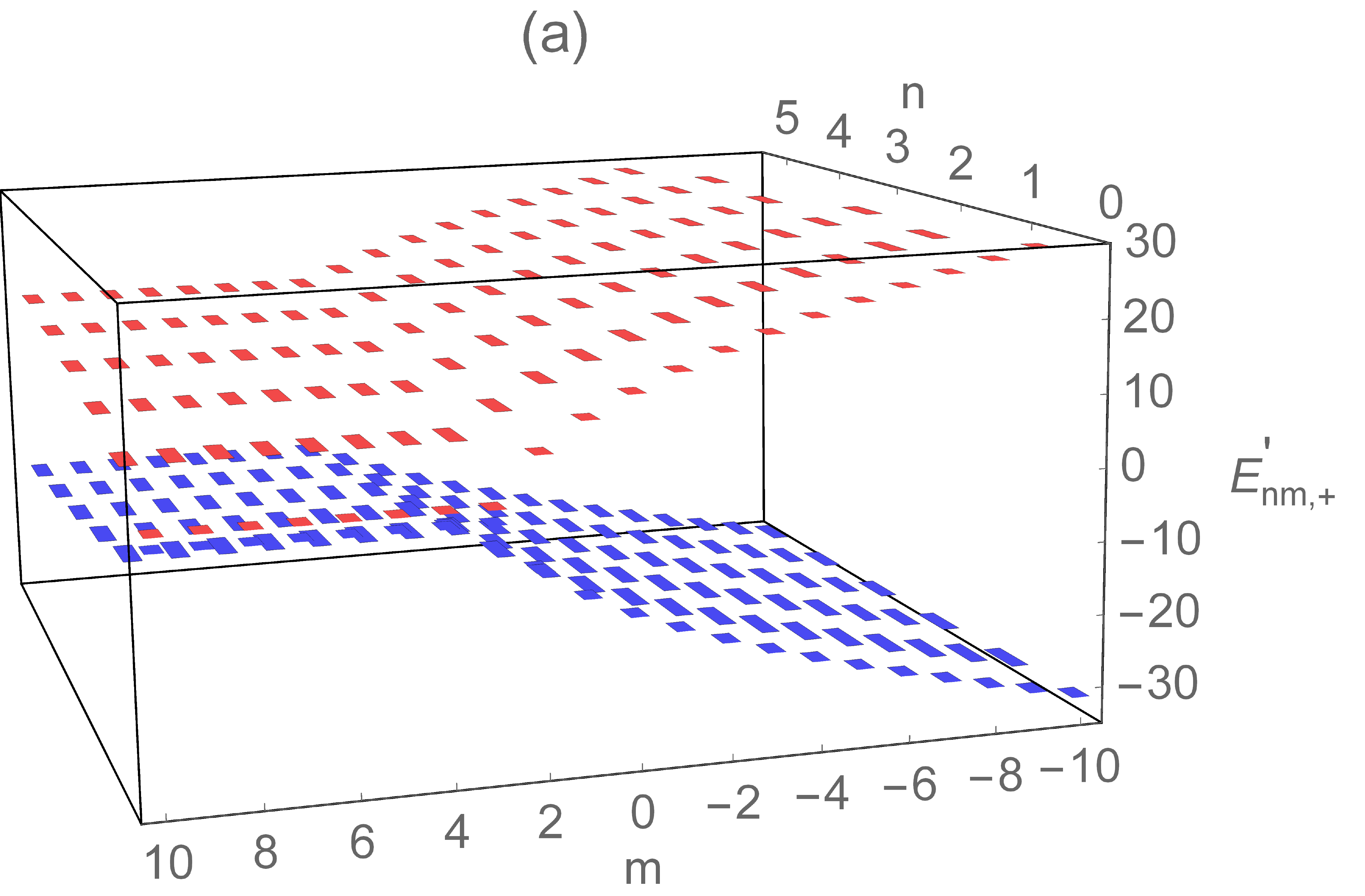}\vspace{0.3cm}
\includegraphics[scale=0.19]{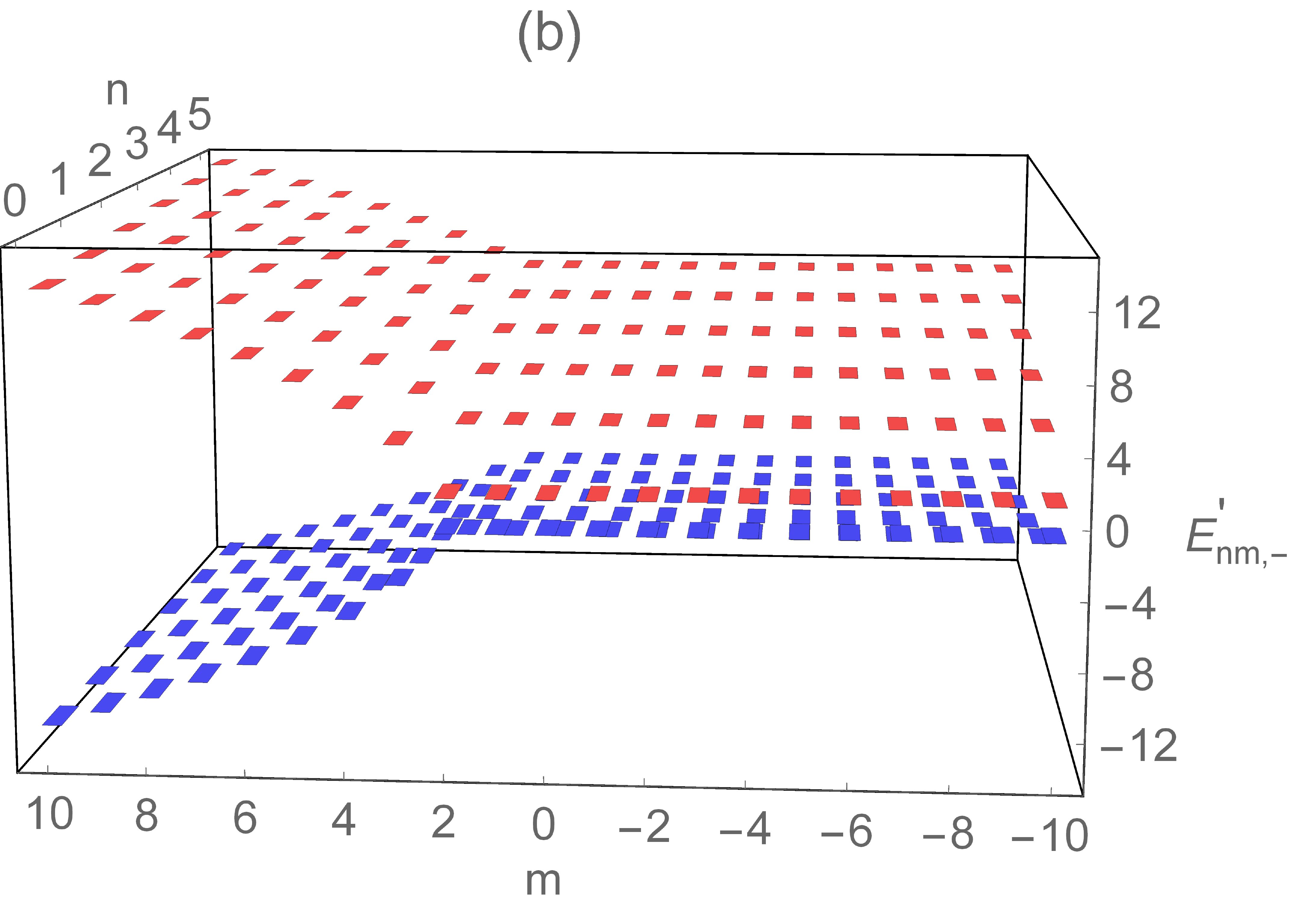}
\caption{(Color online)
Sketch of $E^{\prime}_{nm,+}$ (Eq. (\ref{Enma0})) in Panel (a) and $E^{\prime}_{nm,-}$ (Eq. (\ref{Enma0_2})) in Panel (b) as a function of $n$ and $m$ for the fixed parameters $a=0.1$, $\alpha=0.5$, $\phi=3$, $\omega=6$ and $\omega_{c}=7$.}
\label{Fig3D_Enm_a0}
\end{figure}

In the particular case when the system is not under rotation, the effective angular momentum $L_{\pm}$ in Eqs. (\ref{maef}) and (\ref{maef_2}) no longer depends on energy. In this way, for the Eq. (\ref{rd}), the condition (\ref{cond}) becomes
\begin{equation}
\frac{1}{2}\left( 1+{\frac{\left\vert \ell_{+} \right\vert }{\alpha }}
\right) -{\frac{\varepsilon^{\prime }_{+}}{4M\Omega_{+} }=-n},\label{cond2}
\end{equation}
with
\begin{equation}
\varepsilon^{\prime}_{+} =E^{\prime2}-M^{2}+2M\Omega_{+} \left( 1+\frac{\ell_{+}}{%
\alpha }\right),
\end{equation}
where $\ell_{+}$ is given in Eq. (\ref{maef}). 
By resolving (\ref{cond2}) to $E^{\prime}$, we obtain
\begin{equation}
E^{\prime}_{nm,+}=\pm \sqrt{2M\Omega_{+} \left(2n+\frac{1}{\alpha }\left( \left\vert \ell_{+} \right\vert -\ell_{+}\right) \right) +M^{2}}. \label{Enma0}
\end{equation}

\begin{figure}[!b]
\centering
\includegraphics[scale=0.3]{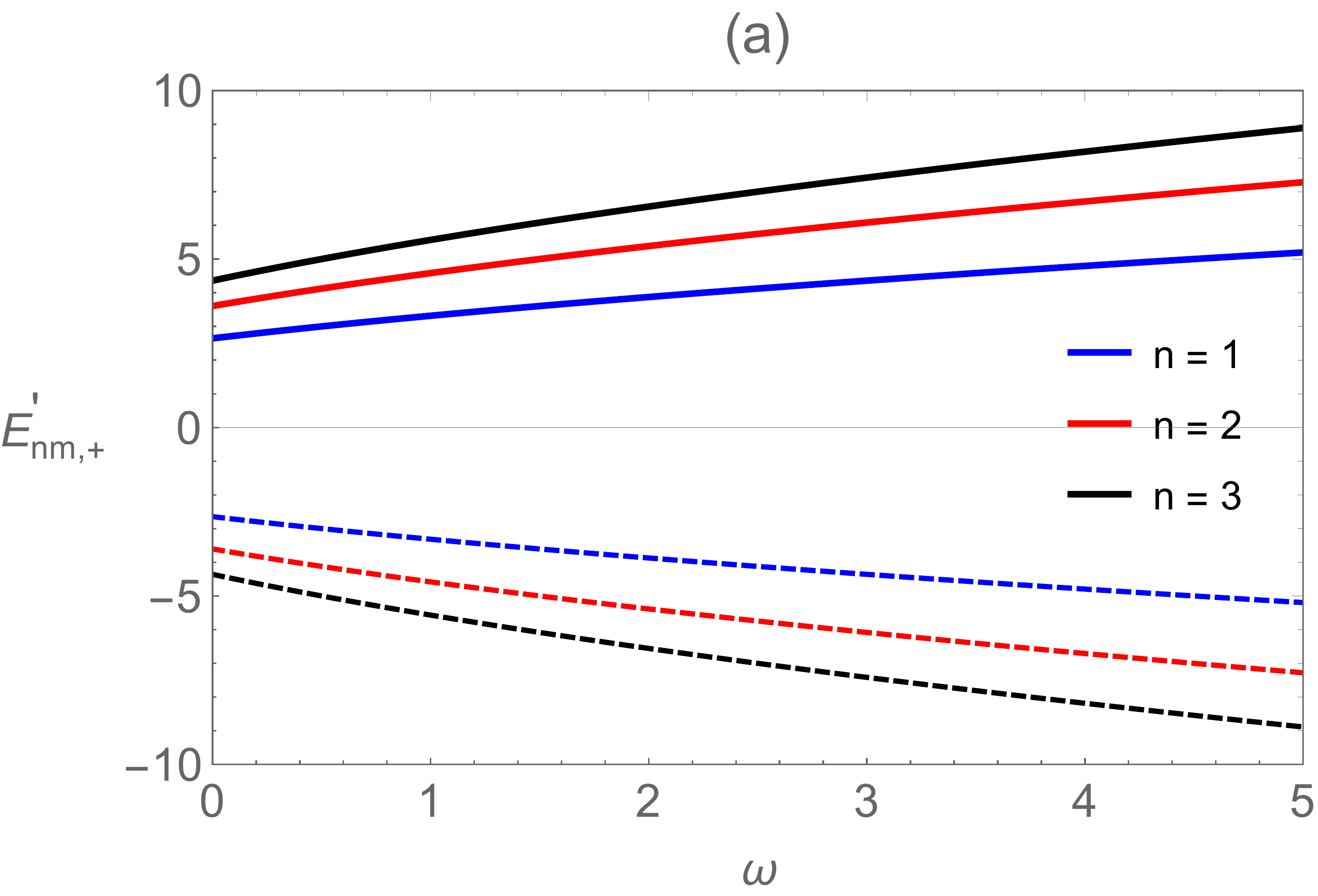}\vspace{0.3cm}
\includegraphics[scale=0.3]{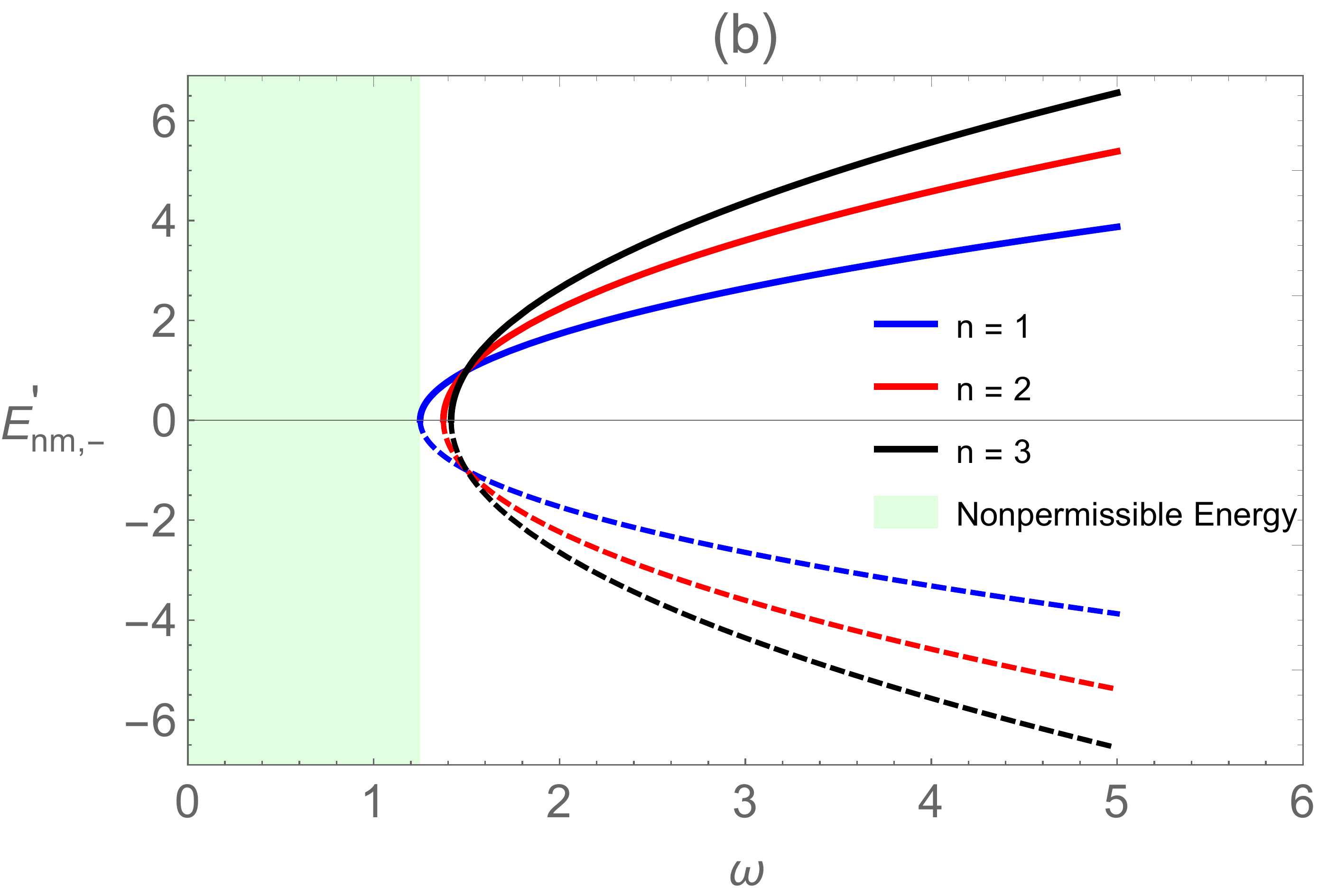}
\caption{(Color online)
Sketch of $E^{\prime}_{nm,+}$ (Eq. (\ref{Enma0})) and $E^{\prime}_{nm,-}$ (Eq. (\ref{Enma0_2})) as a function of $\omega$ for the fixed parameters $\alpha=0.1$, $\phi=3$ and $\omega_{c}=3$. In Panel (a), $m=3$ and in Panel (b), $m=1$.}
\label{Fig2D_Enm_w_a0}
\end{figure}
Similarly, when the system is not rotating, we can also find the energy related to $g(r)$, from Eqs. (\ref{rd_2}) and (\ref{maef_2}):
\begin{equation}
E^{\prime}_{nm,-}=\pm \sqrt{2M\Omega_{-} \left(2n+\frac{1}{\alpha }\left( \left\vert \ell_{-} \right\vert -\ell_{-}\right) \right) +M^{2}}. \label{Enma0_2}
\end{equation}
We see that the energies (\ref{Enma0}) and (\ref{Enma0_2}) depend on all the physical parameters involved. In Fig. (\ref{Fig3D_Enm_a0})(a), we show the profile of $E^{\prime}_{nm,+}$ as a function of $n$ and $m$. Clearly, we see that the particle and antiparticle energies increase when $n$ is increased. Furthermore, we also observe that the largest variations in energy occur in states with $m<0$. However, when we investigate the profile of $E^{\prime}_{nm,-}$, this effect is manifested in the states with $m>0$ (Fig. \ref{Fig3D_Enm_a0}(b)). An important question that should be mentioned is that when we study the spectrum (\ref{Enma0_2}) as a function of $n$ and $m$, we find that for certain parameter values we obtain $E_{nm,-}^{\prime}=\pm M$, which do not belong to the set of eigenvalues of the Eq. (\ref{eav}) and, consequently, in the present case. This can be checked with the parameters $\alpha=0.2$, $\phi=4$, $\omega=1$, $\omega_{c}=2$. By making $\alpha=0.7$, $\phi=10$, $\omega=1$ and $\omega_{c}=9$ in Eq. (\ref{Enma0}), the energy spectrum presents a similar behaviour to the states with $m>3$ showed in Fig. \ref{Fig3D_Enm_a0}(a), for any value of $m$. We also study the profiles of $E^{\prime}_{nm,+}$ and $E^{\prime}_{nm,-}$ as a function of $\omega$ (Fig. \ref{Fig2D_Enm_w_a0}). The profile of $E^{\prime}_{nm,+}$ in Fig. \ref{Fig2D_Enm_w_a0}(a) is similar to the one of $E_{nm,+}^{>}$ shown in Fig. \ref{Fig2D_P_ExFosci_Wc0}. This similarity is partially broken when we look at the profile of $E^{\prime}_{nm,+}$, where we observe an interval of $\omega$ resulting in nonpermissible energies as well as a frequency value $\omega=1.5$ occurring an inversion between the energy states (Fig. \ref{Fig2D_Enm_w_a0}(b)).

From the above results, it is possible to estimate the phenomenological energy scale of the model. For this to be accomplished, we shall restore the constants $c$, $G$, and $\hbar$. We can write the parameters $\alpha$ and $a$ as $\alpha=1-4G\mu/c^2$ and $a=(4GJ/c^3)^{1/2}$. The parameter $a$ has units of length. According to the literature, we can look for bounds involving the cosmic string tension,  $ G\mu/c^{2}$ \cite{Jung_2020_JCAP}.  Several works are dealing with this issue nowadays, having as the main goal the search for cosmological signatures of cosmic strings. Then, observational techniques related to cosmic microwave background (CMB) \cite{PhysRevD.93.123503} and gravitational waves (GW) \cite{PhysRevD.97.123505} are employed.
 Following Ref. \cite{EPJC_74_2972_2014}, current observations limit $G\mu/c^2 < 7.36 \times 10^{-7}.$  It implies that, in the cosmological setting, $\alpha$ must be near to $1$. 
We can also set up the parameters concerning the rotation.
For instance, following Ref. \cite{EPJC(2016)76512}, and adopting an angular momentum $J \approx 10^{47}$ $kg. m^2. s^{-1}$, we obtain $a \approx 10^6$ $m$. 
Besides, the subject of quantifying the intergalactic and the interstellar magnetic fields  also has attracted attention.
For example, from   Refs \cite{EPJC(2016)76512,doi:10.1146/annurev-astro-091916-055221}, we find the magnetic field of the intergalactic medium is $B \approx 10^{-10}$ $T$.
After recovering $c, \hbar$ and $G$, Eq. (\ref{cp2}) can be written as
\begin{equation}
E_{nm,+}^{\left( <\right)}=-E_{1}\pm E_{2},\label{cp2_SI}
\end{equation}
with
\begin{align}
&E_{1}=\frac{2M\Omega_{+} a c}{\alpha },\\
&E_{2}=\frac{1}{\alpha }\notag \\
&\times
\sqrt{4M^{2}\Omega_{+}^{2}a^{2}c^{2}+4M\Omega_{+} \hbar c^{2} \alpha \left( n\alpha -\ell_+ \right) +\alpha^{2}M^{2}c^{4}}. 
\end{align}
In Table I, we show some numerical values for the energies $E_1$ and $E_2$ of Eq. (\ref{cp2_SI}).
Only for $a=10^7$ $m$ and $a=10^6$ $m$  is possible to obtain $E_1$ with a numeric value comparable with $E_2$, with both terms having the same order of magnitude. We can notice that $E_1$ is highly sensitive to changes in the parameter $a$, while $E_2$ does not suffer expressive modifications in several cases if we adopt five significant digits.
If we consider small values for $a$, like $a=1$ or $0.01$  $m$, then $E_1$ reaches the scale of condensed matter physics \cite{PhysRevMaterials.3.024005}. 
\begin{table}[!h]
\centering
 \begin{tabular}{||c | c | c | c |  c ||} 
 \hline
 $a$ (m) \,\, &  B (T) \,\, & \,\, $\alpha$ & \,\, $E_1$ (eV) & \,\, $E_2$ (eV) $\times 10^4$ \\ [2.5ex] 
 \hline\hline
  $10^7$ & $10^{-10}$ & $0.99$ & $60.524 \times 10^4$ & $79.222$ \\
 \hline
 $10^6$ \,\, & $10^{-10}$ \,\, & \,\, $0.99$ & \,\,  $6.0523 \times 10^4$  & \,\, $51.475$  \,\, \\
 \hline
 $10^5$ \,\, & $10^{-10}$ \,\, & \,\, $0.99$ & \,\, $6.0523 \times 10^3$  & \,\, $51.121 $ \,\, \\
 \hline
$10^4$ \,\, & $10^{-10}$ \,\, & \,\, $0.99$ & \,\, $6.0523 \times 10^2$ & \,\, $51.118 $ \,\, \\
 \hline
 $10^2$ \,\, & $10^{-10}$ \,\, & \,\, $0.99$ & \,\, $6.0523$ & \,\, $51.118$  \,\, \\
 \hline
 $1$ \,\, & $10^{-10}$ \,\, & \,\, $0.99$ & \,\, $6.0523 \times 10^{-2}$ & \,\, $51.118 $ \,\, \\
 \hline
 $0.01$ \,\, & $10^{-10}$ \,\, & \,\, $0.99$ & \,\, $6.0523 \times 10^{-4}$ & \,\, $51.118 $ \,\, \\
 \hline
 $0.01$ \,\, & $10^{-6}$ \,\, & \,\, $0.99$ & \,\, $6.0523$ & \,\, $51.118$ \,\, \\
 [1ex] 
 \hline
\end{tabular}
\label{table_energies}
\caption{Some numerical values for the energies in Eq. (\ref{cp2}), for the setting $n=m=1$, $\phi=\omega=0$. The quantity $E1$ refers to the first term of (\ref{cp2}), while $E2$ refers to the second term of the same expression; $B$ indicates the magnetic field (in units of Tesla).}
\end{table}

\section{Connection with Condensed Matter Physics}\label{SecIV}

A curious feature involving the energy spectrum we have obtained in the previous section is related to some analogies with condensed matter physics.  Then, in this section, we explore these similarities. It is known that graphene physics has been working as a bridge between condensed matter physics and relativistic quantum mechanics because, in the low-energy approximation, it is possible to consider that the charge carriers in such material obey an effective Dirac equation for massless particles \cite{RevModPhys.81.109}. This subject has inspired other developments, both in the search for the fabrication of novel materials, as well as new theoretical predictions.
Similarly to graphene, the materials known as topological insulators also have some features involving analogies with the Dirac theory. Topological insulators have attracted expressive attention, due to their exotic properties. A remarkable characteristic of such materials is the fact they have an insulating gap in the bulk, while the edge or surface states are gapless \cite{RevModPhys.83.1057}. In this context, we can take a look at the model introduced by Bernevig, Hughes and Zhang, namely, the BHZ model, describing a type of two-dimensional topological insulator \cite{Bernevig1757,RevModPhys.83.1057}.
In the BHZ model, the bulk energy spectrum is given by
\begin{equation}
E_{\pm}=\epsilon(k) \pm \sqrt{A^2(k_x^2 +k_y^2)+M^2(k)},
\label{bhz_1}
\end{equation}
with
\begin{equation}
\epsilon(k)=C-D(k_x^2+k_y^2), 
\hspace{0.4cm}
M(k)=M-B(k_x^2+k_y^2).   
\end{equation}
The signs $(\pm)$ in Eq. (\ref{bhz_1}) refers to the energies of electron-like and hole-like bands.
In these expressions, $A$, $B$, $C$ and $D$ are material parameters, $k_x, k_y$ are the components of the momentum $\mathbf{k}$ and $M$ is the mass parameter.
The dispersion relation of BHZ model is similar to the energies (\ref{cp2}) and (\ref{cpm1_2}). The first term of these expressions, which has a dependence on the rotation parameter, is the analog of the term $\epsilon(k)$ in Eq.  (\ref{bhz_1}).

In the BHZ model, the mass parameter is a function of $k_x$ and $k_y$. In the energies (\ref{cp2}) and (\ref{cpm1_2}), for example, we have a similar term inside the square root involving the square of the mass, namely, $M^{2}[4a^2 \Omega_{\pm}^2+\alpha^2]$. Thus, in both models, the effective mass parameter in the square root depends on properties of the system: in the BHZ model, it is a function of the momentum components $k_x$ and $k_y$, while in our model, the mass parameter shows up jointly with the rotation and curvature parameters. In this way, the presence of the rotation makes possible the comparison between the energy spectrum of the BHZ model and the energy of the Dirac oscillator derived here.
The energy in the BHZ model can either increase or decrease, with respect to the ``quantum numbers" $k_x$ and $k_y$.

In the usual Dirac oscillator, i.e, without noninertial effects and other interactions, the energy increases as a function of the quantum number $n$.
On the other hand, we have shown here that the combination of noninertial and topological effects as well as the magnetic interactions can modify the hierarchy of the energy levels in some cases. For instance, in the cases depicted in Figs. \ref{Fig2DP_E_W_osc}(a) \ref{Fig2DP_E_W_osc}(b), we observe the existence of a region where the energy decreases in relation to $n$ (an inversion of the energy levels). Thus, in our model, the energy can either increases or decreases, as a function of the quantum number $n$. Also, while in the model described here we verify an inversion of the energy levels, in the case of topological insulators, in some situations, it is possible to observe an inversion of the electron-like and hole-like energy bands \cite{RevModPhys.83.1057,RevModPhys.88.021004}.

Another relevant point in this comparison refers to the energy gap. While the gap between the energy bands of electrons and holes in the context of Condensed Matter systems depends on the materials parameters, in our case, the gap between particle and antiparticle states depends on parameters related to rotation, curvature, and magnetic field and the Aharonov-Bohm potential.

\section{Conclusions}
\label{SecV}

In this article, we have investigated how the combination of non-inertial and topological effects modify the energy levels of the two-dimensional Dirac oscillator in the presence of a uniform magnetic field and the Aharonov-Bohm effect. We have obtained the Dirac equation describing the Dirac oscillator spinning cosmic string spacetime in the presence of such a configuration of magnetic fields. The resulting motion equation revealed that the effective angular moment of the particle explicitly depends on its energy. This fact is a physical implication due to the rotation of the spacetime in which the particle lives. To avoid the singular potential coming from $\mathbf{\nabla} \times \mathbf{A}$, we have solved the eigenvalue equation in the $r \neq 0$ region. We have derived the eigenfunctions and the corresponding energy spectrum for the problem. The eigenfunctions are given in terms of the confluent hypergeometric function. Concerning the energy spectrum, because of its explicit dependence on the effective angular moment of the particle, the eigenvalues problem provides two different situations, corresponding to the cases $L_{\pm}>0$ and $L_{\pm}<0$. The presence of the
rotation produces the appearance of exotic characteristics in
 the energy spectrum. The combined effect involving rotation, curvature and external fields reveals that the energy spectrum can present different behaviors. An immediate study showed that the projection element of spin leads to the energies (\ref{cp1}), (\ref{cp2}), (\ref{cpm1_2}) and (\ref{cpm2_2}). This particularity revealed that the energies (\ref{cp1}) and (\ref{cpm2_2}) depend only on the frequency of the oscillator $\omega$, the magnetic field $B$ and the quantum number $n$ and, therefore, exhibiting symmetric shape (Fig. 	\ref{Fig2D_P_ExFosci_Wc0}). Because of the similarity between the energies (\ref{cp2}) and (\ref{cpm1_2}), we prefer to investigate only (\ref{cpm1_2}). The reason for this is that after several analyses of (\ref{cp2}), we have found profiles similar to (\ref{cpm1_2}), changing only the intensity of the energies. Our investigation was centered on the combined effects involving rotation, curvature, magnetic field, and the Aharonov-Bohm flux. We have studied the energy profile (\ref{cpm1_2}) in several aspects. In some cases, the shape of the energy spectrum can be drastically modified by changing the physical parameters of the system. This occurred when we analyzed its profile as a function of $\omega$ (Figs. \ref{Fig2DP_E_W_osc}(a)-(b)), where we have verified an inversion between states and a range of $\omega$ with prohibited energies. The parameters can be controlled in such a way that no energy state is absent (Fig. \ref{Fig2DP_E_W_osc}(c)). In all cases analyzed, we identified the intervals of the parameters that result in forbidden energies by the colored parts highlighted in the figures that presented these characteristics. We have also found inversion between states in the energy profile as a function of the cyclonic frequency $\omega_{c}$ (Figs. \ref{Fig2DP_E_Wciclo}(a)-(b)), as a function of rotation (Fig. \ref{Fig2DP_Erot}(b)) and as a function of curvature (Fig. \ref{Fig2DP_Ealpha}(c)). In all cases where the inversion effect between states occurred, $|E_{nm,-}|$ decreases when $n$ is increased. In other cases, $|E_{nm,-}|$ increases as $n$ increases. In the particular case without the presence of rotation, we recover the symmetrical form of the energy levels as it occurs in the Landau relativistic quantization (Fig. \ref{Fig3D_Enm_a0}). However, even in the absence of rotation, we verified the presence of prohibited energies when we investigated their profile as a function of the frequency $\omega$ of the oscillator. This effect occurs due the fact the effective frequency $\Omega_{+}$ (Eq. (\ref{Omeff})) and $\Omega_{-}$ (Eq. (\ref{Omeff_2})) depends on the spin component. 

The complexity inherent to the model addressed in this article has allowed us (in some situations) to notice some similarities between the energies found here and the energy spectrum of a Topological Insulator. This similarity between such models can serve as motivation to study models in relativistic quantum mechanics in rotating frames with applications in condensed matter physics.

\section*{Acknowledgments}
We would like to acknowledge the anonymous reviewer for the relevant
criticism and suggestions.
The authors also thank Manoel M. Ferreira Jr. and Rodolfo Casana for their valuable discussions.
This work was partially supported by the Brazilian agencies CAPES, CNPq and
FAPEMA. Edilberto O. Silva acknowledges CNPq Grants 427214/2016-5 and 307203/2019-0, and
FAPEMA Grants 01852/14 and 01202/16. This study was financed in part by the Coordena\c{c}\~{a}o de
Aperfei\c{c}oamento de Pessoal de N\'{\i}vel Superior - Brasil (CAPES) -
Finance Code 001. M\'{a}rcio M. Cunha acknowledges CAPES
Grant 88887.358036/2019-00.

\bibliographystyle{apsrev4-2}
\bibliography{References}

\end{document}